\documentclass{article}
\usepackage[utf8]{inputenc}
\usepackage[margin=1.0in]{geometry}
\usepackage{amsmath,graphicx,url}
\usepackage{color}
\usepackage{booktabs}
\usepackage{hyperref}

\title{DIFFERENTIABLE WORLD SYNTHESIZER-BASED \\ NEURAL VOCODER WITH APPLICATION TO END-TO-END AUDIO STYLE TRANSFER}
\author{Shahan Nercessian}
\date{iZotope, Inc.}

\begin{document}

\maketitle

\begin{sloppy}
\begin{abstract}
In this paper, we propose a differentiable WORLD synthesizer and demonstrate its use in end-to-end audio style transfer tasks such as (singing) voice conversion and the DDSP timbre transfer task.  Accordingly, our baseline differentiable synthesizer has no model parameters, yet it yields adequate synthesis quality.  We can extend the baseline synthesizer by appending lightweight black-box postnets which apply further processing to the baseline output in order to improve fidelity.  An alternative differentiable approach considers extraction of the source excitation spectrum directly, which can improve naturalness albeit for a narrower class of style transfer applications.  The acoustic feature parameterization used by our approaches has the added benefit that it naturally disentangles pitch and timbral information so that they can be modeled separately.  Moreover, as there exists a robust means of estimating these acoustic features from monophonic audio sources, it allows for parameter loss terms to be added to an end-to-end objective function, which can help convergence and/or further stabilize (adversarial) training.
A revised version of this preprint has been accepted to the 154th AES Convention.  To cite this work, please refer to the AES manuscript available at \href{https://www.aes.org/e-lib/browse.cfm?elib=22073}{https://www.aes.org/e-lib/browse.cfm?elib=22073}.
\end{abstract}

\section{Introduction}
\label{sec:intro}
Vocoders aim to solve the inverse problem of reconstructing audio from a compressed acoustic feature representation \cite{Tacotron}.  Efficient, high-quality vocoders have become a basic requirement for neural audio synthesis, with applications in text-to-speech \cite{Tacotron}, audio style transfer \cite{Ddsp}, singing voice synthesis \cite{SingingSynthesis}, speech coding \cite{LPCNet}, etc.  In such cases, the vocoder tends to act as a synthesis back-end in some greater system, synthesizing audio from features that have been estimated and/or transformed by some feature processing front-end \cite{Tacotron}.

Core signal processing vocoders for speech synthesis (though extendable to monophonic sources in general) include STRAIGHT and WORLD \cite{WORLD}.  The WORLD vocoder is particularly notable, as it pairs a hand-designed spectral feature analysis procedure with a dedicated synthesis algorithm mapping its representation back to a phase coherent time-domain signal.  As with many innovations in the field, these methods have been largely replaced by their deep learning counterparts of late.  Neural vocoders such as WaveNet \cite{WaveNet}, WaveRNN \cite{WaveRnn}, and LPCNet \cite{LPCNet} are neural auto-regressive generative models which model audio at the sample level, and have improved realism in many speech applications and related target domains.  Despite their quality gains, their auto-regressive nature tends to bottleneck throughput.  Later innovations such as MelGAN have avoided the need for auto-regression by formulating an efficient architecture framed as a sheer upsampling of feature representations at frame rate to samples at audio rate \cite{MelGAN}.  Its use of adversarial training, particularly the inclusion of a deep feature loss defined in terms of discriminator feature maps, improves system fidelity without needing to explicitly parameterize the output probability distributions of audio samples.  The feature representation used by most neural vocoders tends to be the log Mel spectrogram \cite{Tacotron}, though this is not a strict requirement.  It should be noted that these vocoders generally do not have any built-in guarantees of phase coherence or the preservation a desired pitch contour in their outputs, which is particularly crucial in musical applications. As such, the singing voice synthesis community, may still lean heavily on the WORLD vocoder, despite the many advances in neural vocoding \cite{SingingSynthesis, WganSing}.

More recently, a complementary class of vocoder algorithms including source-filter models \cite{Nsf2} and differentiable digital signal processing (DDSP) \cite{Ddsp} explicitly model monophonic source production in a parametric manner, hearkening back to the core signal processing approaches of the past.  Such techniques have been successfully extended to speech applications, either as standalone vocoders \cite{HooliGan, Ddspeech} or as part of an end-to-end system \cite{Shahan3}.  However, the main disadvantage of these approaches is that they lack a deterministic analysis procedure to accurately extract vocoder parameters directly from source audio.  This is to say that one must still learn a mapping which translates audio or a given audio representation to their respective vocoder parameters.

In this work, we formulate a differentiable WORLD synthesizer and illustrate its use in end-to-end style transfer applications.  The premise of this work comes from the observation that the WORLD vocoder has largely been neglected since the advent of neural vocoders, although, for monophonic sources, it has desirable features which still cannot be matched entirely by the current "state-of-the-art."  Specifically, it contains an explicit mechanism for preserving a desired pitch contour in its outputs, and is paired with analysis procedure to accurately extract acoustic features which may be treated as "ground truth."  This means that as long as we adhere to the WORLD feature representation (and/or compressed derivatives described in this paper), the resulting differentiable synthesizer can yield decent synthesis quality without any model parameters, while leveraging a feature representation which naturally disentangles pitch and timbre so that they can be modeled separately.  Additionally, we can append lightweight, non-autoregressive black-box networks to the end of the synthesizer in order to post-filter the raw synthesizer output in an effort to further improve synthesis quality.  Alternatively, we also propose a differentiable framework which does not synthesize an excitation signal, and rather manipulates the formants of a source signal by extracting the spectrum of its excitation and imposing a new spectral envelope according to a given task.

The remainder of this paper is organized as follows.  Section \ref{sec:vocoder} introduces our proposed differentiable WORLD synthesizer and its variants.  Section \ref{sec:e2e} extends the application of the proposed neural vocoder to different audio style transfer tasks.  Section \ref{sec:results} illustrates experimental results, while Section \ref{sec:conclusions} draws conclusions.

\section{Differentiable WORLD synthesizer-based neural vocoder}
\label{sec:vocoder}

\subsection{Baseline differentiable WORLD synthesizer}
\label{ssec:baseline}
WORLD analysis yields frame-level fundamental frequency $f_0$, spectral envelope $sp$ (as a power spectrum), and aperiodicity ratio $ap$ acoustic features from which a synthesized waveform of adequate quality can be readily calculated \cite{WORLD}.  We notionally follow the processing carried out by the original, non-differentiable WORLD synthesizer implementation, making some reasonable compromises in order to implement the synthesizer using vectorized differentiable operations which better accommodate efficient backpropogation using standard deep learning frameworks.

Our differentiable WORLD synthesizer implementation consists of the synthesis of harmonic and noise signal components.  In harmonic synthesis, we begin by synthesizing a pulse train excitation per the detected fundamental frequency contour.  A rudimentary implementation of this involves synthesis of a time-varying sine tone followed by hard-clipping and rising edge detection.  This will cause aliasing at fundamental frequencies for which the sampling rate $f_s$ is not an integer multiple (to this end, the non-differentiable WORLD synthesizer efficiently reduces aliasing via fractionally-shifted pulses).  Alternatively, we consider an alias-free design via a summation of sinusoids.  The number of harmonics is chosen to span the entire frequency range assuming a minimum "floor" fundamental frequency $f_{min}$ (set to 71 Hz per the WORLD vocoder's default settings).  Operating at a sampling rate of 22050 Hz, this results in $K=155$ sinusoids characterizing a fundamental and its $K-1$ harmonics.  Upon interpolating $f_0$ to audio rate (denoted as $\tilde{f_0}$), the pulse train $e_h$ is synthesized via
\begin{equation}
\label{eqn:pulse}
e_h = \sum_{k=1}^Kc_k\sin(2\pi k\tilde{f_0}t)
\end{equation}
where time-varying harmonic amplitudes $c_k$ are deterministically computed in order to both mask harmonics depending on whether their corresponding instantaneous frequency is above the Nyquist rate, as well as to normalize the energy of each resulting pulse.

Given $sp$ and $ap$, the pulse train excitation is filtered via
\begin{equation}
\label{eqn:harm}
h = \mathcal{F}^{-1}\left[(1-ap)\odot\sqrt{sp}\odot\mathcal{F}(e_h)\right]
\end{equation}
where $\mathcal{F}$ denotes an $N$-point, Hann-windowed short-time Fourier transform (STFT) assuming some notional hop size of the synthesizer (here we default to $N/4$ samples), yielding the harmonic component $h$.  Noise synthesis involves generation of a random noise excitation $e_n$ (sampled from $\mathcal{N}(0, 1)$).  The noise excitation is filtered via
\begin{equation}
\label{eqn:nz}
n = \mathcal{F}^{-1}\left[ap\odot\sqrt{sp}\odot\mathcal{F}(e_n)\right]
\end{equation}
yielding the noise component $n$.  It is worth noting that $ap$ is set to unity for all time-frequency bins for which an unvoiced frame is detected.  This ensures that during unvoiced frames, no harmonic signal is present and that noise shaping is purely a function of $sp$.  The output of our baseline differentiable WORLD synthesizer is then given by
\begin{equation}
\label{eqn:baseline}
y_0 = g_hh+g_nn
\end{equation}
where $g_h$ and $g_n$ are user-specified gain parameters that can be varied at inference time, but are set to unity during model training.
\subsection{Parameter compression/decompression}
\label{ssec:compression}
It should be noted that at a target sampling rate of 22050 Hz, WORLD analysis/synthesis involves a 1024-point STFT, giving rise to 513-element spectral envelope and aperiodicity ratio vectors at each time step (these feature dimensions double at standard audio rate). In the context of neural audio synthesis, where such a vocoder would be used as a back-end synthesizer, estimation of this many acoustic parameters by a front-end feature transformation network may be impractical, and has consistently been shown to be unnecessary.  Accordingly, we consider a sound, differentiable compression/decompression procedure with strong analogues to existing neural vocoder methodologies.  The compression scheme is useful for determining targets for an acoustic feature loss, whereas the decompression scheme is implemented as part of the differentiable synthesizer to deterministically transform compressed representations so that they are interpretable to the synthesizer.  We define a so-called WORLD log Mel spectrogram as
\begin{equation}
\label{eqn:world_mels_compress}
s = \log_{10}(\mathbf{M}\sqrt{sp}+\epsilon)
\end{equation}
where $\mathbf{M}$ is the Mel (or similarly defined log-spaced) basis matrix used to compute an $M$-band Mel spectrogram from an $N$-point STFT, and $\epsilon$ is a small constant used for numerical stability.  The WORLD log Mel spectrogram is analogous to a standard log Mel spectrogram (and thus, we consider $M=80$ as in \cite{Tacotron}), except that it is largely void of pitch information.  This is an important distinction between feature representations, and a key advantage of the proposed one, as in many style transfer contexts, this means that feature transformations front-ends only need to infer timbral information so long as pitch information can be derived from the source content.  The decompressed approximation is then given by
\begin{equation}
\label{eqn:world_mels_decompress}
sp^\dagger = \left[\mathbf{M}^\dagger_0(10^s-\epsilon)\right]^2
\end{equation}
where $\mathbf{M}^\dagger_0 = \max(\mathbf{M}^\dagger, 0)$ and $\mathbf{M}^\dagger$ denotes the pseudo-inverse of $\mathbf{M}$.  Note that we can avoid the $\sqrt{(\cdot)^2}$ operations for numerical stability, but include them here for illustrative purposes.  Since spectral envelopes are significantly smoother than raw spectra, the compression procedure reduces feature dimensions significantly, yet impart very little in the way of signal degradation (especially as phase information is captured independently via the excitation signal).  We illustrate the proposed compression/decompression scheme in Figure \ref{fig:spec}, as applied to both a standard magnitude spectrogram and its estimated spectral envelope.  We observe that indeed, the WORLD log Mel spectogram is highly analogous to the standard log Mel spectrogram that is ubiquitious in neural audio synthesis applications, except that we have proportionally smoothed out a considerable amount of pitch information from its feature representation so that we only characterize signal timbre.

\begin{figure*}[ht]
  \begin{minipage}{.33\textwidth}
	\centering
  \centerline{\includegraphics[width=0.95\columnwidth]{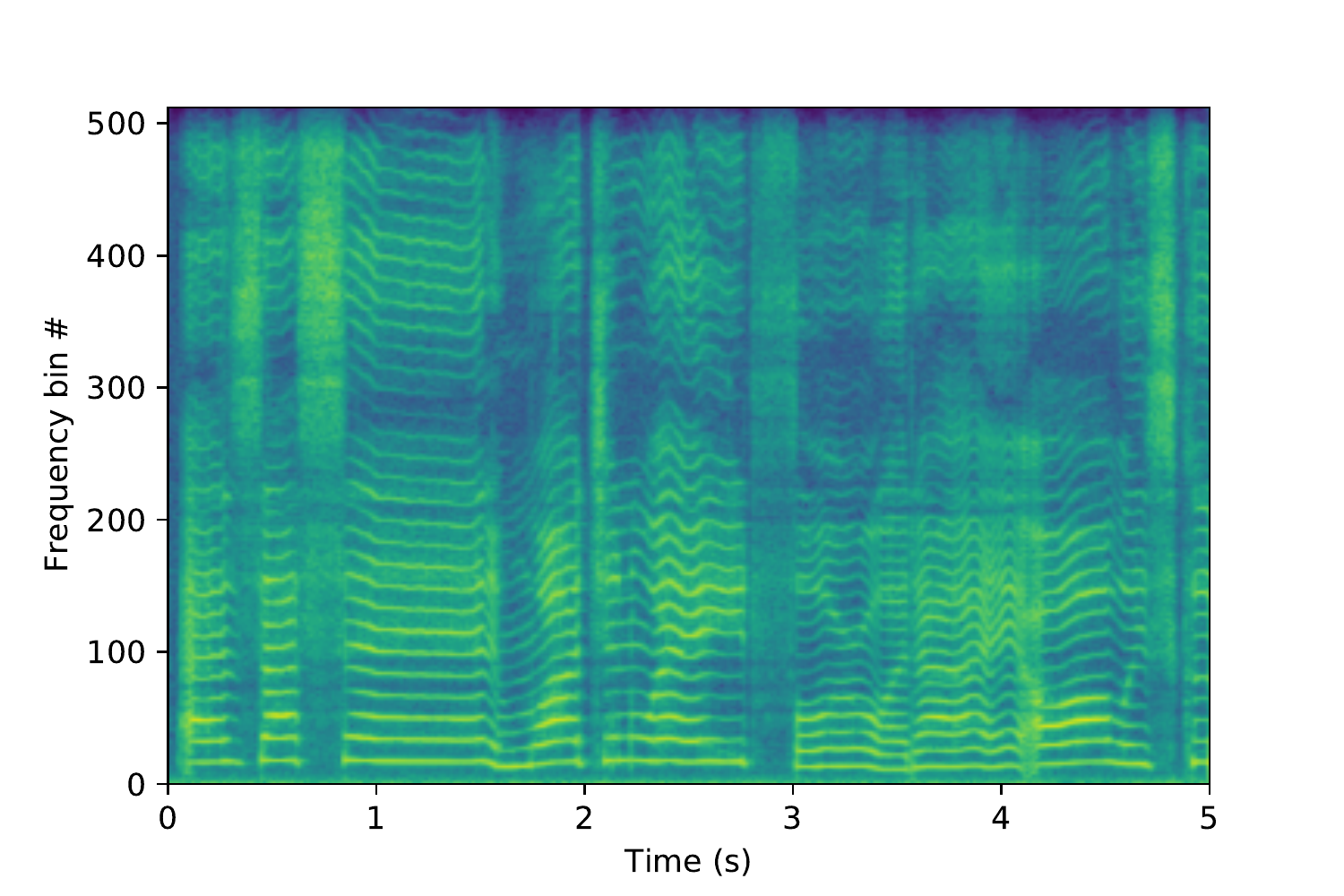}}
	\vspace{0.25em}(a)
  \end{minipage}
  \begin{minipage}{.33\textwidth}
	\centering
  \centerline{\includegraphics[width=0.95\columnwidth]{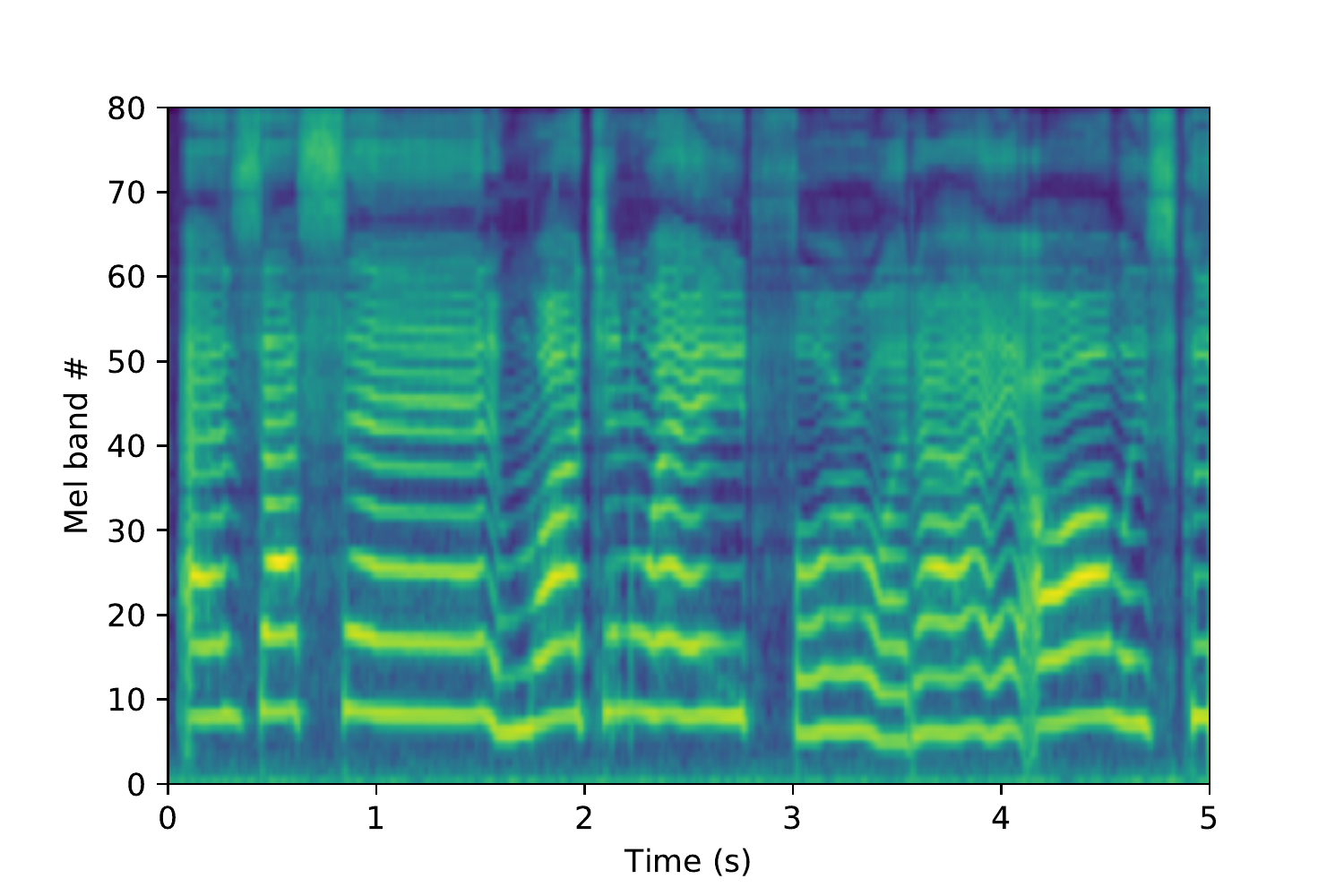}}
	\vspace{0.25em}(b)
  \end{minipage}
  \begin{minipage}{.33\textwidth}
	\centering
  \centerline{\includegraphics[width=0.95\columnwidth]{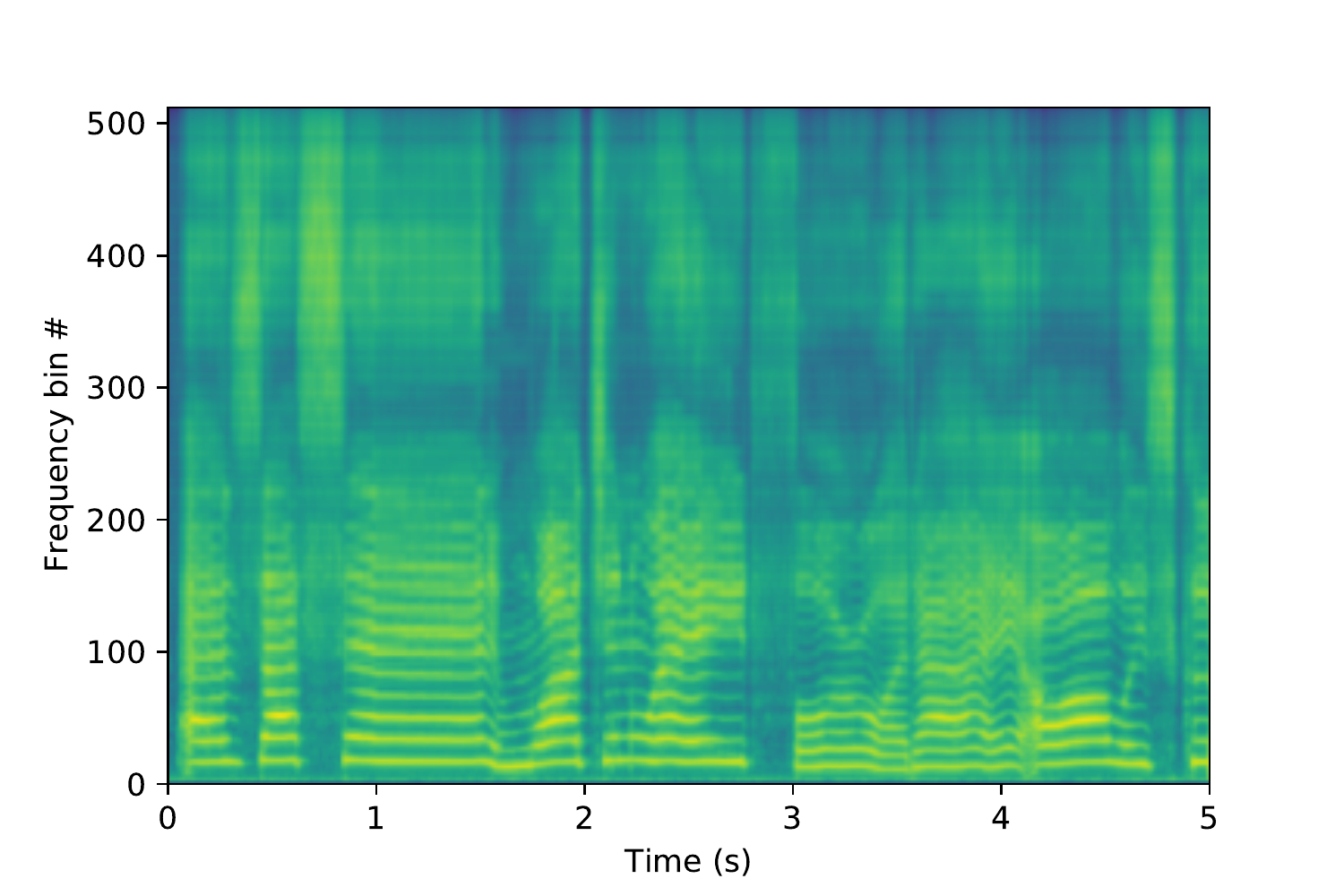}}
	\vspace{0.25em}(c)
  \end{minipage}

\begin{minipage}{.33\textwidth}
	\centering
  \centerline{\includegraphics[width=0.95\columnwidth]{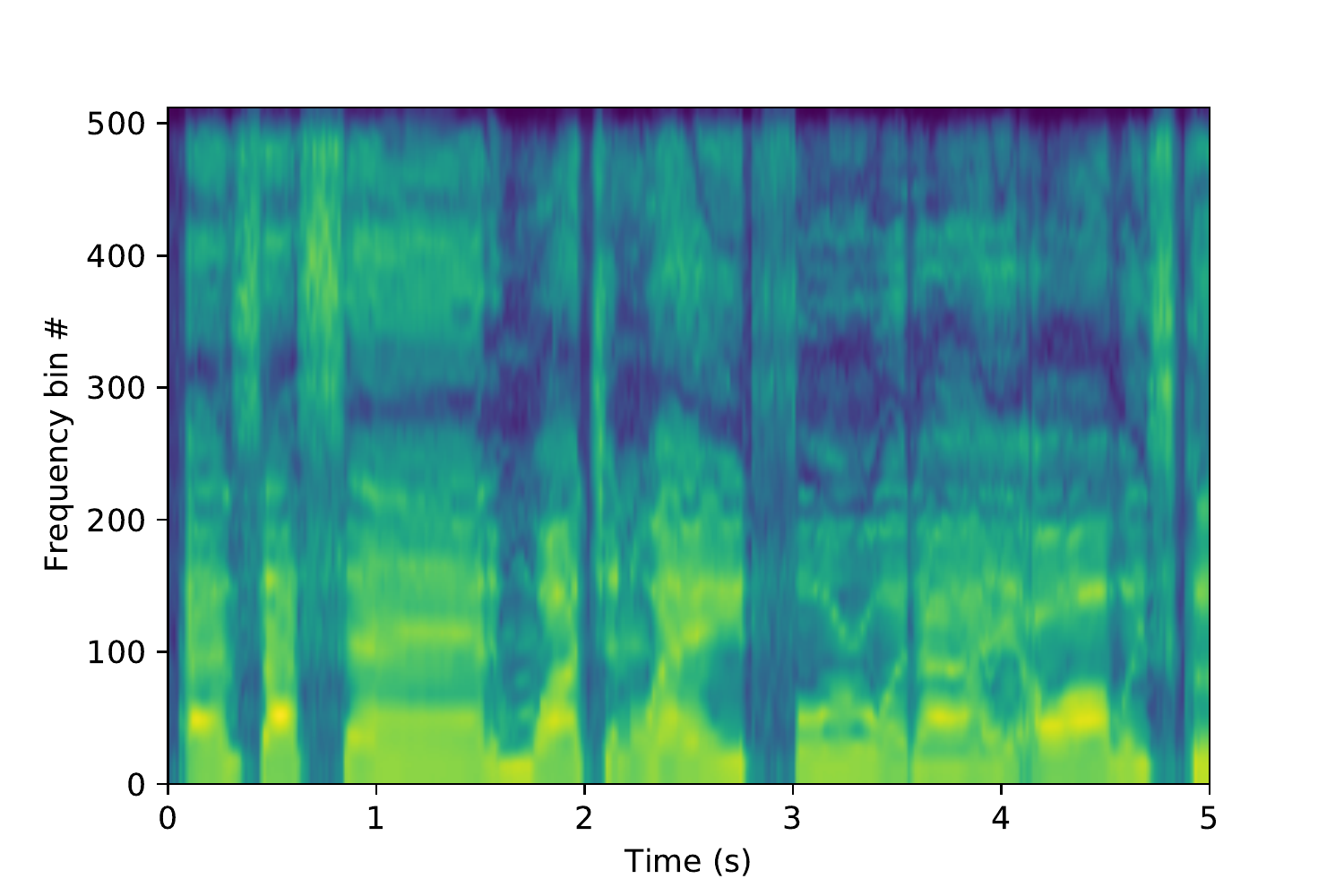}}
	\vspace{0.25em}(d)
  \end{minipage}
  \begin{minipage}{.33\textwidth}
	\centering
  \centerline{\includegraphics[width=0.95\columnwidth]{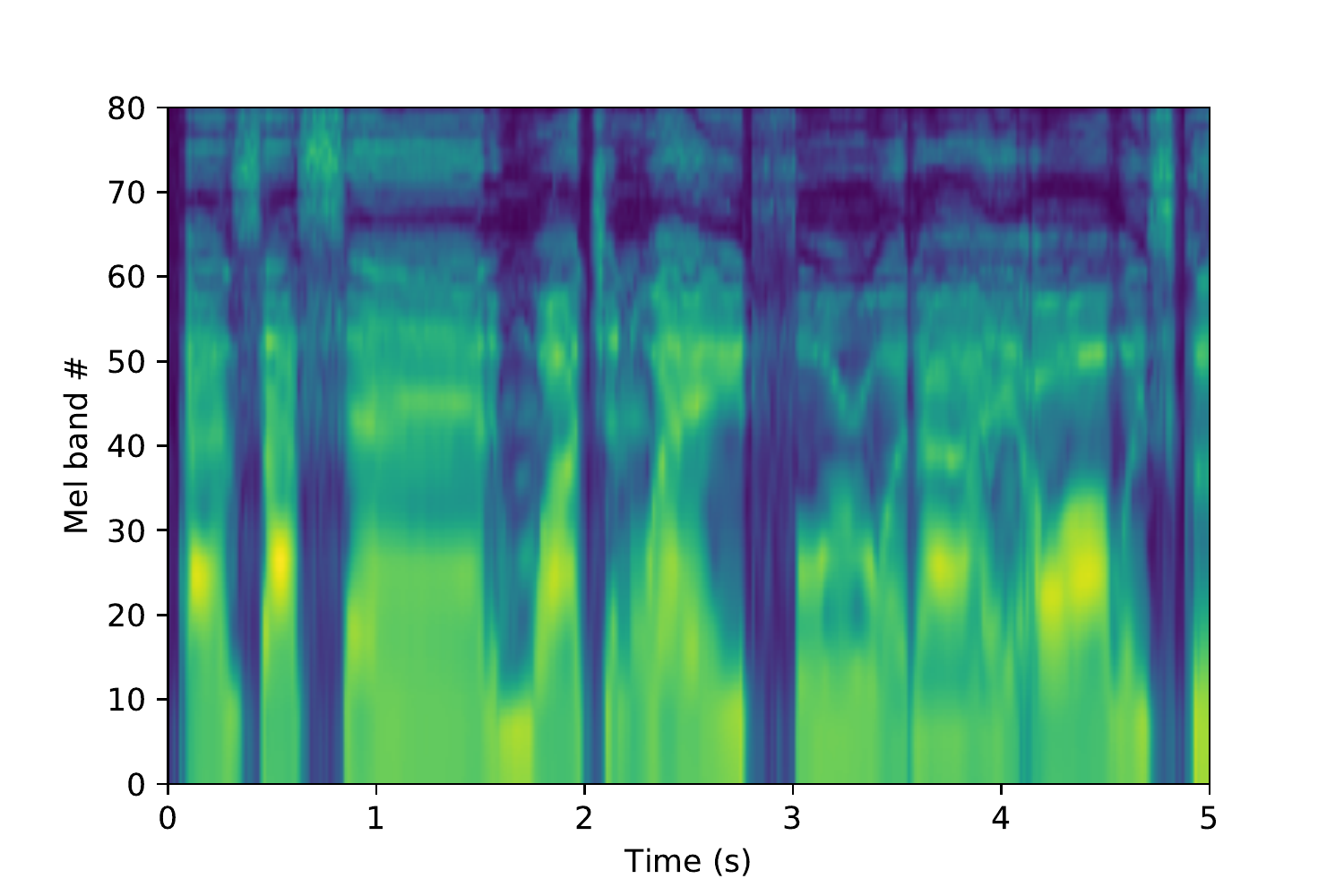}}
	\vspace{0.25em}(e)
  \end{minipage}
  \begin{minipage}{.33\textwidth}
	\centering
  \centerline{\includegraphics[width=0.95\columnwidth]{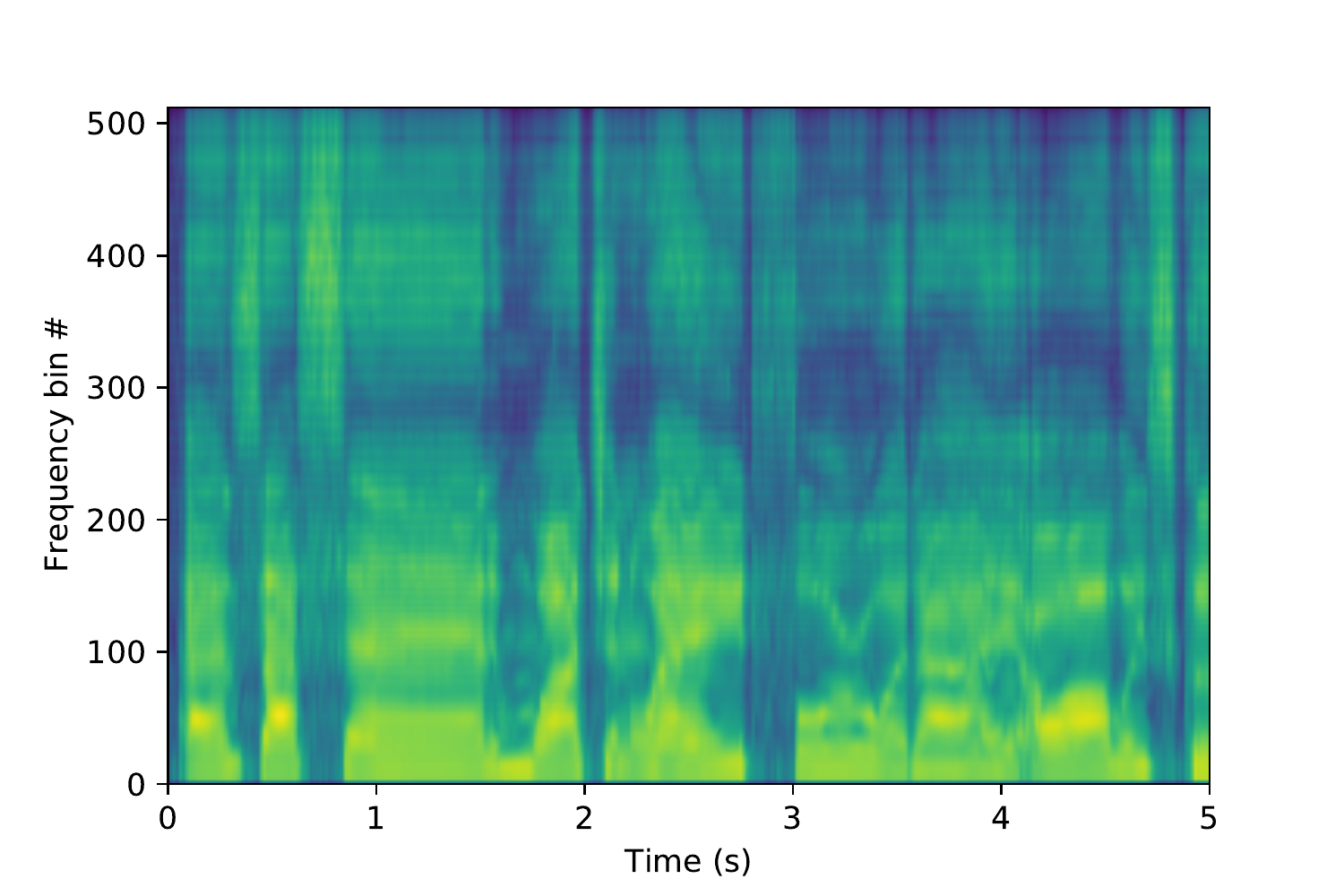}}
	\vspace{0.25em}(f)
  \end{minipage}

\caption{\it (a) (Log) magnitude spectrogram, (b) log Mel spectrogram, (c) log Mel spectrogram inversion, (d) (log) WORLD spectral envelope, (e) corresponding WORLD log Mel spectrogram,  and (f) WORLD log Mel spectrogram inversion.}
\label{fig:spec}
\end{figure*}

Compression/decompression of $ap$ involves downsampling/upsampling onto regular frequency grids by means of linear interpolation, respectively, which retains aperiodicity ratio bounds to remain in [0, 1] by design.  To this end, we empirically found that the use of 16 compressed components was sufficient for modeling aperiodicity ratio while introducing minimal additional artifacts to the synthesis, and denote the compressed and decompressed aperiodicity ratios as $a$ and $ap^\dagger$ respectively.  Interestingly, the WORLD feature representation itself adds implicit bias for the classical voice conversion scenario in which case we only care to replace speaker identity, as the aperiodicity component need not be inferred by the feature transformation front-end.  In this case, we can use the source aperiodicity directly and do not need to perform its respective compression/decompression scheme (i.e. $ap^\dagger=ap$).

\subsection{Comparison to the DDSP synthesizer}
\label{ssec:hns}
The synthesizer used in the original DDSP paper \cite{Ddsp} also considers generation of harmonic and noise signal components, and is thus related to our proposed baseline synthesizer.  It offers a similar level of expressivity, but is parameterized differently.  There are practical considerations which we believe make the parameterization of the differentiable WORLD synthesizer proposed here to be more fruitful.  Specifically, a useful byproduct of the proposed scheme is its acoustic features can be deterministically estimated from ground truth source audio \cite{WORLD}, whereas, at the time of writing this document, they can only be learned via an analysis-by-synthesis procedure in the case of the DDSP synthesizer.  We posit that the former is more desirable, as this allows us to include an acoustic feature loss term into the training objective, which we intuit would help stabilize training.  Consequently, by pairing the synthesizer to our proposed WORLD feature representation, the baseline synthesizer does not require any additional model parameters of its own, whereas the DDSP synthesizer requires a transformation network to translate an arbitrary acoustic feature representation into the parameters that it understands.  Admittedly, in the context of the DDSP timbre transfer (DDSP-TT) task \cite{Ddsp}, our method would still require a feature decoder to transform pitch and loudness measurements into their corresponding compressed WORLD synthesizer parameters.

\subsection{Extensions using black-box residual postnets}
\label{ssec:postnet}
We believe that our differentiable WORLD synthesizer, though not perfect on its own, provides significant implicit bias to the task of neural vocoding, allowing us to forego the need for auto-regression, and to generally reduce neural vocoder complexity relative to current standards.  To this end, we propose appending black-box post-filtering networks which further transform the raw output from the differentiable WORLD synthesizer.  The motivation of such postnets are two-fold:  we would like for 1) the postnet to use its additional degrees of freedom to improve the fidelity of the vocoder output, and 2) to learn a deep prior of the target domain at the sample level.  The latter point is particularly crucial in transformative neural audio synthesis applications, because a significant source of system degradation arises when inferred acoustic feature distributions deviate from their target domain(s).  Accordingly, it is seen that most waveform-based neural vocoders trained on speech, for example, often "vocalize" source acoustic features regardless of their domain, inherently performing their own form of "style transfer," so to speak.  While it is not strictly necessary, we consider postnets which add a residual signal to the differentiable WORLD synthesizer outputs, both to provide further end-user tunability and to leverage the faster training offered by residual learning.  Under this paradigm, given a postnet $P$, the post-filtered output becomes
\begin{equation}
\label{eqn:postnet}
\begin{split}
y_d & = g_0y_0+g_PP(y_0) \\
  & = g_0y_0+g_Py_P
\end{split}
\end{equation}
where $g_0$ and $g_P$ are user-specified gain parameters set to unity during training. To model $P$, we can consider, amongst other time-domain architectural candidates, the temporal convolutional network as in \cite{HifiGan} or the SEANET network as in \cite{SeaNet}.  The latter is particularly fruitful, as it has been shown to be effective while operating several times faster than real-time.  Each layer in $P$ can optionally be locally conditioned on acoustic features as in \cite{HifiGan2}.  This allows the filtering operation to adapt itself as a function of a more global context vector.  Specifically, we consider the Mel spectrogram of $y_0$ upsampled to the system sampling rate using linear interpolation as local conditioning for each layer of the postnet, which we can compute in an online fashion.  For the case of the SEANET architecture, we must downsample this conditioning information in order to match the dilation factor at each layer.

Finally, as in \cite{Ddsp, Shahan3}, we can optionally learn an impulse response $w$ to apply to $y_d$, which can either characterize reverb or simply act as a final learned global equalization filter.  The filter $w$ is specifically designed to be causal, and additionally with $w(0)=0$.  As such, the output of the vocoder becomes
\begin{equation}
\label{eqn:vocoder}
\begin{split}
y & = g_dy_d + g_w\left(y_d*w\right) \\
 & = g_dy_d + g_wy_w
\end{split}
\end{equation}
where $g_d$ and $g_w$ are user-specified gain parameters set to unity during training. The resulting vocoder, now with free model parameters, can be trained on ground truth acoustic features extracted from target audio, or incorporated within a larger end-to-end neural audio synthesis system, in which case acoustic features are provided from some transformative front-end.
 
\subsection{Alternate source excitation method}
\label{ssec:alternate}
Alternatively, we can consider a "synthesis-free" method to signal manipulation and/or reconstruction which extracts the excitation directly from source audio.  The excitation can be estimated from a source signal $x$ via
\begin{equation}
\label{eqn:excitation}
e = \mathcal{F}^{-1}\left[\frac{1}{\sqrt{sp}}\odot\mathcal{F}(x)\right]
\end{equation}
The signal can be reconstructed via
\begin{equation}
\label{eqn:synthesis}
y = \mathcal{F}^{-1}\left[\sqrt{sp}\odot\mathcal{F}(e)\right]
\end{equation}
Note that in practice, 1) either the $sp$ or $sp^\dagger$ form can be considered (depending on the specific application), 2) there is no immediate need to extract the intermediate excitation time-domain signal as its spectrum is sufficient, 3) the construction of $y$ does not need $f_0$ or $ap$, and 4) formant manipulation is achieved when the spectral envelopes in (\ref{eqn:excitation}) and (\ref{eqn:synthesis}) are made different from one another.  Accordingly, given the spectral envelope $sp$ extracted from audio signal $x$ and some derived manipulation $\hat{sp}$, a transformed waveform is determined by

\begin{equation}
\label{eqn:formant}
y = \mathcal{F}^{-1}\left[\sqrt{\frac{\hat{sp}^{(\dagger)}}{sp^{(\dagger)}}}\odot\mathcal{F}(x)\right]
\end{equation}
where $(\dagger)$ serves to denote that either the raw or decompressed representation could be used to this end.  Naturally, with $\hat{sp}=sp$ (or alternatively, $\hat{sp}^\dagger=sp^\dagger$) we have $y = x$.  Therefore, the advantage of this alternate approach is that, as the excitation is derived directly from source audio as opposed being synthesized, an identity operation is possible (unlike the synthesis-based approach).  However, the practical use of this approach is limited to formant transformations, and would require other external mechanisms for additional audio manipulations (i.e. formant-preserving pitch shifting \cite{Lent}).  As such, this rigidity is likely acceptable and can provide potentially more realistic results for a task like singing voice conversion (SVC), but is less applicable to other applications necessitating a vocoder, such as the DDSP-TT task, text-to-speech, singing voice synthesis, etc.  To this end, $\hat{sp}$ could be derived via an interpolation of $sp$ onto a (potentially irregularly spaced) set of grid points in frequency.  Similarly, $\hat{sp}^\dagger$ could be derived via a feature transformation network (yielding a non-linear, time-varying formant transformation), as described in the following section.

\section{Application to audio style transfer}
\label{sec:e2e}

In this section, we integrate the proposed methodologies into autoencoder-based audio style transfer models for SVC and the DDSP timbre task, as notionally illustrated in Figure \ref{fig:nn_arch}.  An encoder extracts source-independent representations from audio, while a decoder learns to map them to suitable acoustic features for the given task.  Lastly, a differentiable WORLD-based vocoder generates output waveforms from the inferred acoustic features.  With the advent of end-to-end training, model weights are refined by comparing ground truth and inferred waveforms, including those of a suitable vocoder postnet if one is enabled.  In the case of SVC, we are primarily concerned with inferring spectral envelopes representative of the target vocalist(s) which we are attempting to model, and can use the aperiodicity ratio of the input signal directly as needed.  In the case of the DDSP-TT task, we additionally need to infer signal aperiodicity, as the time-varying noise profile (capturing breathiness, bows/plucks, etc.) contributes to the aesthetics of a particular monophonic instrument (this is analogous to the independent inference of harmonic and noise parameters in \cite{Ddsp}).  In both cases, we are primarily modeling timbre, and assume that the detected source pitch contour of a performance can be used directly and/or deterministically adapted (i.e. suitable octave shifts can be determined as per \cite{Shahan2}).  In other applications, such as singing voice synthesis, text-to-speech, etc., an explicit prosody model would be necessary in order to infer the output pitch contour, but such applications are not covered here.

\begin{figure*}[t]
\center
\includegraphics[width=6.5in]{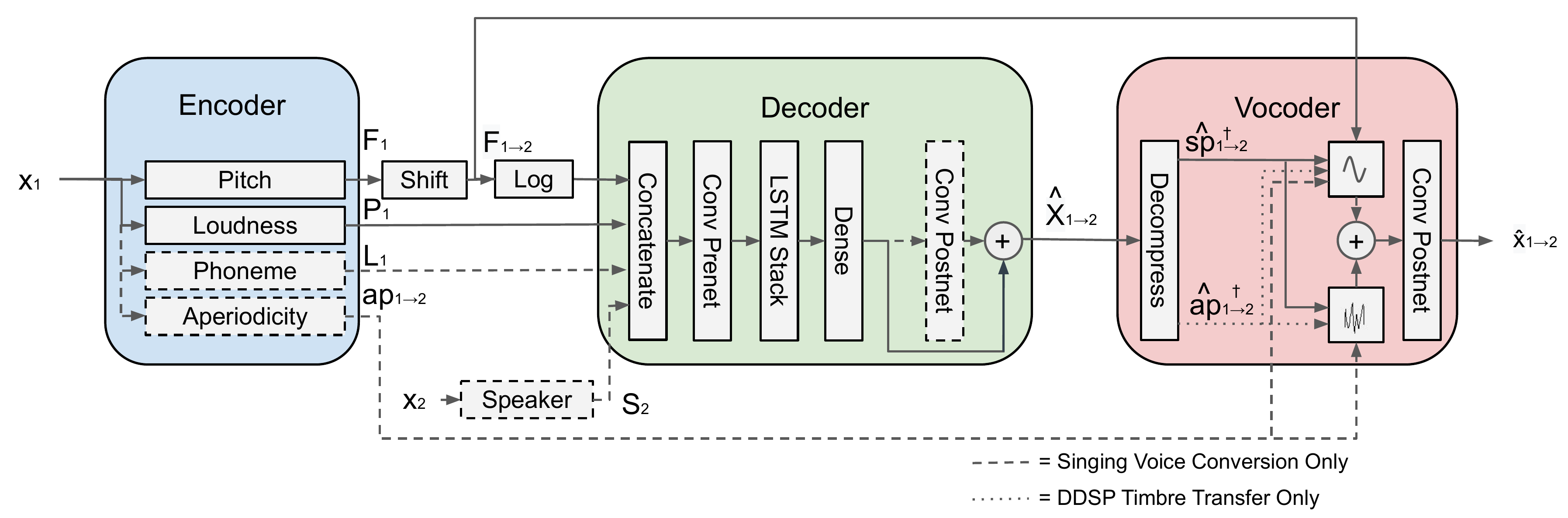}
\caption{\label{fig:nn_arch}{\it Block diagram outlining the audio style transfer methods considered here.  Components exclusive to SVC are contained in dashed regions.  Components exclusive to the DDSP-TT task are contained in dotted regions.  For SVC, the output waveform can alternatively be computed using our source excitation variant (not pictured here).}}
\vspace*{-5pt}
\end{figure*}

\subsection{Singing voice conversion}
\label{ssec:svc}

We adapt an SVC architecture used in some of our previous works, replacing the vocoder module with the differentiable WORLD-based neural vocoder variants proposed here.  We provide a brief overview of the system here, and refer readers to our earlier works for further details in \cite{Shahan3, Shahan1, Shahan2}.

We consider performances (waveforms) $x_1$ and $x_2$ from source and target vocalists, respectively, and begin by extracting a vocalist-independent encoding of $x_1$.  Source loudness $L_1$ is extracted deterministically.  We leverage the tonality-gated $f_0$ contour $F_1$ extracted during WORLD analysis, and at inference time, can shift it (to the nearest octave) considering the median pitch of the source of target vocalists, yielding the target pitch contour $F_{1\rightarrow2}$. Phonetic information $P_1$ is captured in the form of a phonetic posteriorgram inferred using a bidirectional LSTM architecture trained on the TIMIT dataset \cite{Timit}.  The encoding is given by the concatenation of loudness, octave-adjusted pitch, and phonetic information $Z_{1\rightarrow2} = \left[ L_1, \log{F_{1\rightarrow2}}, P_1 \right]$. During WORLD analysis, we can also extract the aperiodicity ratio of $x_1$, but it need not be an input to the system decoder.  As mentioned previously, since it may be passed directly to the vocoder module which generates the converted waveform, we denote it as ${ap}_{1\rightarrow2}$.

The decoder then maps $C_{1\rightarrow2} = \left[ L_1, \log{F_{1\rightarrow2}}, P_1, S_2 \right]$ to acoustic features $\hat{X}_{1\rightarrow2} = \hat{s}_{1\rightarrow2}$, where $S_2$ is a speaker embedding extracted from the performance of the target vocalist waveform $x_2$.  During training, we have $x_1=x_2$, and specifically, source and target performance waveforms reflect the same vocalist (i.e. the system is trained as an autoencoder).  However, at inference time, the source and target vocalists can vary, and this, in turn with the vocalist-independent encoding, becomes the mechanism for enabling zero-shot voice conversion in practice if the system were trained over a plethora of vocalists.

To generate the output waveform, we begin by applying feature decompression to $\hat{X}_{1\rightarrow2}$, which in this case simply yields $\hat{sp}^\dagger_{1\rightarrow2}$. We can now leverage any of the vocoder variants in Section {\ref{sec:vocoder}}.  If we intend on using a variant which synthesizes a new excitation signal, we feed $\hat{sp}^\dagger_{1\rightarrow2}$ and ${ap}_{1\rightarrow2}$ along with $F_{1\rightarrow2}$ in order for the differentiable synthesizer (with potential postnet) to generate the output waveform $\hat{x}_{1\rightarrow2}$.  If we opt for the method in Section \ref{ssec:alternate}, we need $\hat{sp}^\dagger_{1\rightarrow2}$ and a properly pitch-shifted input waveform from which an excitation spectrum can be extracted.

\subsection{DDSP timbre transfer}
\label{ssec:ddsp}
Similarly, we can adapt the methodology to the DDSP-TT task, in which case the encoder again extracts loudness and adjusted pitch information from source audio (yielding the encoding $C_{1\rightarrow2} = \left[ L_1, \log{F_{1\rightarrow2}}\right]$) and a decoder maps $C_{1\rightarrow2}$ into suitable synthesizer parameters $\hat{X}_{1\rightarrow2}$.  The decoder applies a 1x1 convolution layer to $C_{1\rightarrow2}$, followed by a single unidirectional LSTM layer (both with 256 units).  A final linear projection layer then yields the raw decoder output, which is split into harmonic and aperiodic components.  A sigmoid activation can be applied to the aperiodicity components so that they remain bounded in [0, 1], which ensures phase coherence. The inferred synthesizer parameters $\hat{X}_{1\rightarrow2} = [\hat{s}_{1\rightarrow2}, \hat{a}_{1\rightarrow2}]$  are fed as inputs to the proposed vocoder.  Feature decompression now yields $\hat{sp}^\dagger_{1\rightarrow2}$ as well as $\hat{ap}^\dagger_{1\rightarrow2}$.  Along with $F_{1\rightarrow2}$, the decomposed features are passed to the differentiable synthesizer which generates the output waveform $\hat{x}_{1\rightarrow2}$.

\subsection{System training}
\label{ssec:system}
Using the methodologies outlined here, audio style transfer systems can be trained in a multitude of ways using a combination of different strategies.  As the class of algorithms described here are trained as an autoencoder, we have $x_1 = x_2 = x_{1\rightarrow2}$, $X_1 = X_2 = X_{1\rightarrow2}$, $F_1=F_2=F_{1\rightarrow2}$, etc. during training unless stated otherwise.  As such, we interchange "1" and "2" subscripts for illustrative purposes while maintaining notational consistency with previously defined variables.

\subsubsection{End-to-end training}
\label{sssec:e2e}
In more conventional training settings, our system decoder and vocoder sub-networks would be trained independently.  The decoder can be trained by defining the mean squared error (MSE) on acoustic features as an objective function, such as
\begin{equation} \label{eqn:mse}
\mathcal{L}_{MSE}(X_{1\rightarrow2}, \hat{X}_{1\rightarrow2}) = \mathbf{E}[|X_{1\rightarrow2}-\hat{X}_{1\rightarrow2}|^2_2]
\end{equation}
Moreover, the vocoder can be trained using ground truth acoustic features $X_1$ extracted from source audio $x_1$.  To this end, a suitable objective function for comparing inferred audio $\hat{x}_1$ to ground truth is the multi-spectrogram loss (MSL), defined as
\begin{equation}
\mathcal{L}_{MSL}(x_1, \hat{x}_1) = \sum_{s=1}^{S} ( \mathbf{E}[|\mathcal{F}_{s}(x_1)-\mathcal{F}_{s}(\hat{x}_1)|_1]~+ \\ \kappa \mathbf{E}[|\log{\mathcal{F}_{s}(x_1)}-\log{\mathcal{F}_{s}(\hat{x}_1)}|_1])
\end{equation}
where $\kappa$ is a hyperparameter (set to 1 in this work), and $\mathcal{F}_s(\cdot)$ denotes the $s$th of $S=6$ magnitude spectrograms, computed with an STFT window size of $2^{5+k}$ samples and 75\% overlap between frames.  Note that in our case, training our vocoder variants independently is only beneficial when we include a black-box postnet, as we generally have no free model parameters otherwise.

A more ideal end-to-end training objective compares audio from the composite style transfer system $\hat{x}_{1\rightarrow2}$ to their corresponding ground truth $x_{1\rightarrow2}$. Accordingly, we can consider the end-to-end objective function
\begin{equation} 
\mathcal{L}_{MSL}(x_{1\rightarrow2}, \hat{x}_{1\rightarrow2}) = \sum_{s=1}^{S} ( \mathbf{E}[|\mathcal{F}_{s}(x_{1\rightarrow2})-\mathcal{F}_{s}(\hat{x}_{1\rightarrow2})|_1]~+ \\ \kappa \mathbf{E}[|\log{\mathcal{F}_{s}(x_{1\rightarrow2})}-\log{\mathcal{F}_{s}(\hat{x}_{1\rightarrow2})}|_1])
\end{equation}
and the parameters of both the decoder and vocoder sub-networks can be optimized jointly according to this objective.

\subsubsection{Adversarial training}
\label{sssec:gan}
Until now, the objective functions defined here reflect the common training methodology whereby model weights are updated in order to minimize the negative log likelihood of its outputs.  To this end, we can include both acoustic feature and waveform loss terms, or
\begin{equation}
\mathcal{L}_{NLL}(X_{1\rightarrow2}, \hat{X}_{1\rightarrow2}, x_{1\rightarrow2}, \hat{x}_{1\rightarrow2}) = \\ \alpha\mathcal{L}_{MSE}(X_{1\rightarrow2}, \hat{X}_{1\rightarrow2}) + \beta\mathcal{L}_{MSL}(x_{1\rightarrow2}, \hat{x}_{1\rightarrow2})
\end{equation} 
where $\alpha$ and $\beta$ are hyperparameters (both set to 1 in this work).  Meanwhile, adversarial training promotes plausible system outputs as determined by a discriminator network \cite{MelGAN}.  Hybrid approaches combining these methodologies have grown increasingly more common \cite{HooliGan, HifiGan, HifiGan2}, and balance average system performance with output signal plausibility. To this end, we adopt the adversarial loss and multi-scale discriminator architecture in \cite{MelGAN}.

We proceed to formulate the adversarial loss for the end-to-end training case, noting that it is entirely possible to define a similar loss when training a vocoder on its own, as in \cite{MelGAN, HooliGan}.  Accordingly, the end-to-end adversarial loss for the generator is
\begin{equation}
\mathcal{L}_{G,ADV}(x_{1\rightarrow2}, \hat{x}_{1\rightarrow2}) = \\ \mu\mathcal{L}_{G,HNG}(\hat{x}_{1\rightarrow2}) + \lambda\mathcal{L}_{DFM}(x_{1\rightarrow2}, \hat{x}_{1\rightarrow2})
\end{equation} 
where the first term is the generator hinge loss, the second term is a deep feature matching loss comparing intermediate discriminator feature maps computed from inferred and target waveforms, and $\mu$ and $\lambda$ are hyperparameters (set to 1 and 10 in this work).  The generator hinge loss is given by
\begin{equation} \label{eqn:ghinge}
\mathcal{L}_{G,HNG}(\hat{x}_{1\rightarrow2}) = \sum_{k=1}^{K}\mathbf{E}[-D_k(\hat{x}_{1\rightarrow2})]
\end{equation}
and aggregates losses across $K=3$ discriminator networks, where each network $D_k$ analyzes $\hat{x}_{1\rightarrow2}$ downsampled by a factor of $2^{k-1}$. The generator deep feature matching loss is given by
\begin{equation} 
\mathcal{L}_{G,DFM}(x_{1\rightarrow2}, \hat{x}_{1\rightarrow2}) = \\ \sum_{k=1}^{K}\sum_{i=1}^{I}\frac{1}{N_i}\mathbf{E}[|D_{k}^{(i)}(x_{1\rightarrow2})-D_{k}^{(i)}(\hat{x}_{1\rightarrow2})|_1]
\end{equation} 
where $D_{k}^{(i)}$ denotes the $i$th layer feature map output and $N_i$ denotes the number of feature maps in layer $i$.  Conversely, the discriminator loss is given by
\begin{equation}
\mathcal{L}_{D,ADV}({x}_{1\rightarrow2},\hat{x}_{1\rightarrow2}) = \mu\sum_{k=1}^{K}(\mathbf{E}[\min(0, 1-D_k(x_{1\rightarrow2}))] \\ +
\mathbf{E}[\min(0, 1+D_k(\hat{x}_{1\rightarrow2}))])
\end{equation}
Lastly, the hybrid objective function for model training of the generator is the sum of the negative log likelihood and adversarial terms, or
\begin{equation}
\mathcal{L}(X_{1\rightarrow2}, \hat{X}_{1\rightarrow2}, x_{1\rightarrow2}, \hat{x}_{1\rightarrow2}) = \\ \mathcal{L}_{NLL}(X_{1\rightarrow2}, \hat{X}_{1\rightarrow2}, x_{1\rightarrow2}, \hat{x}_{1\rightarrow2}) + \mathcal{L}_{G, ADV}(x_{1\rightarrow2}, \hat{x}_{1\rightarrow2})
\end{equation} 

\subsubsection{Constrained-manifold targets for training}
\label{sssec:manifold}
In the specific instance where we use a synthesis variant of the differentiable WORLD vocoder, but have disabled the use of a postnet module, we can redefine the target waveform $x_{1\rightarrow2}$.  Specifically, we can compare the output of our system $\hat{x}_{1\rightarrow2}$ to a new $x_{1\rightarrow2}$ which is synthesized using the baseline WORLD synthesizer with oracle acoustic parameters $X_{1\rightarrow2}$.  This effectively "aligns" targets to live in the constrained manifold of signals that the baseline differentiable WORLD vocoder is capable of synthesizing.  Note that this is rather unique to our vocoder formulation, and is only possible because we can readily estimate WORLD acoustic features from a source audio signal, from which we can adequately produce a synthesized waveform.

\section{Experimental results}
\label{sec:results}

For audio demos, we refer readers to \href{https://sites.google.com/izotope.com/diffworld-audio-demo}{https://sites.google.com/izotope.com/diffworld-audio-demo}.

\subsection{Inversion of WORLD acoustic features}

We begin by demonstrating the capability of our WORLD-based neural vocoder variants to invert WORLD acoustic features, generating monophonic source audio of good quality while preserving pitch, all while being fully differentiable and without necessarily needing free parameters.  To this end, we reconstruct a short singing voice excerpt from acoustic features extracted using the standard WORLD analysis procedure, as illustrated in Figure \ref{fig:inversion}.  Despite some differences, we note that the reconstruction of ground truth WORLD features using the non-differentiable WORLD synthesizer is already quite good.  We observe only perceptually negligible differences between reconstructions using the original non-differentiable synthesizer in Figure \ref{fig:inversion}(b) and with our baseline differentiable counterpart from Section \ref{ssec:baseline} in Figure \ref{fig:inversion}(c).  We can also see that the compression/decompression scheme introduced in Section \ref{ssec:compression} imparts little in the way of additional signal degradation, and similarly, Figure \ref{fig:inversion}(d) resembles Figure \ref{fig:inversion}(b).  If we allow the vocoder to have trainable parameters, as suggested in Section \ref{ssec:postnet}, we can learn a postnet to apply further processing to the model output in Figure \ref{fig:inversion}(e), which can filter feature estimation errors (such as the unvoiced detection towards the end of the clip) and improve realism.  In this case, we have trained the vocoder with a combination of multi-spectrogram and adversarial loss terms using a dataset of the same vocalist (more on this in the following section).  Lastly, it is trivial to show that the source excitation method from Section \ref{ssec:alternate} provides a perfect reconstruction of the signal in this case, as in Figure \ref{fig:inversion}(f).

\begin{figure*}[ht]
  \begin{minipage}{.33\textwidth}
	\centering
  \centerline{\includegraphics[width=0.95\columnwidth]{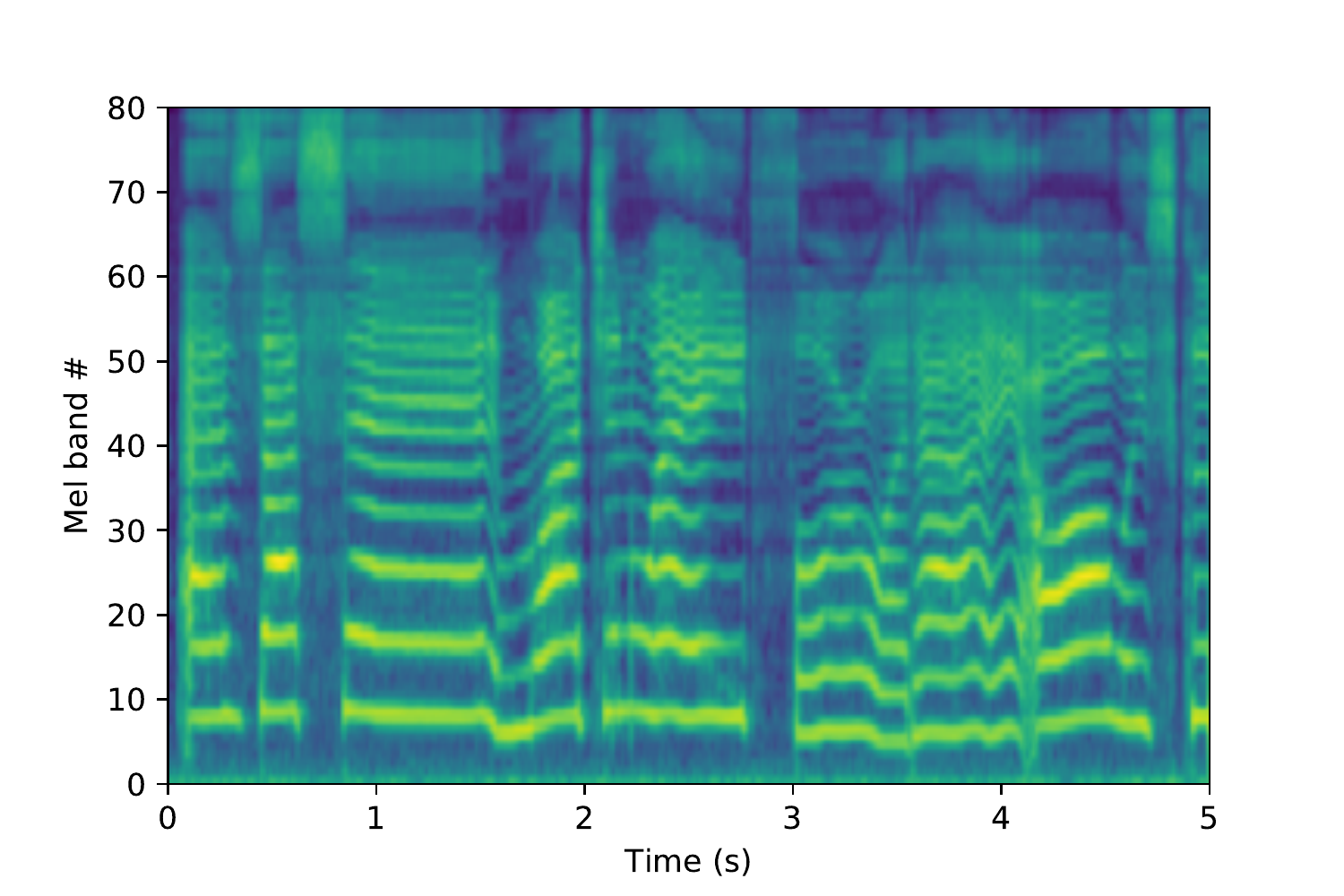}}
	\vspace{0.25em}(a)
  \end{minipage}
  \begin{minipage}{.33\textwidth}
	\centering
  \centerline{\includegraphics[width=0.95\columnwidth]{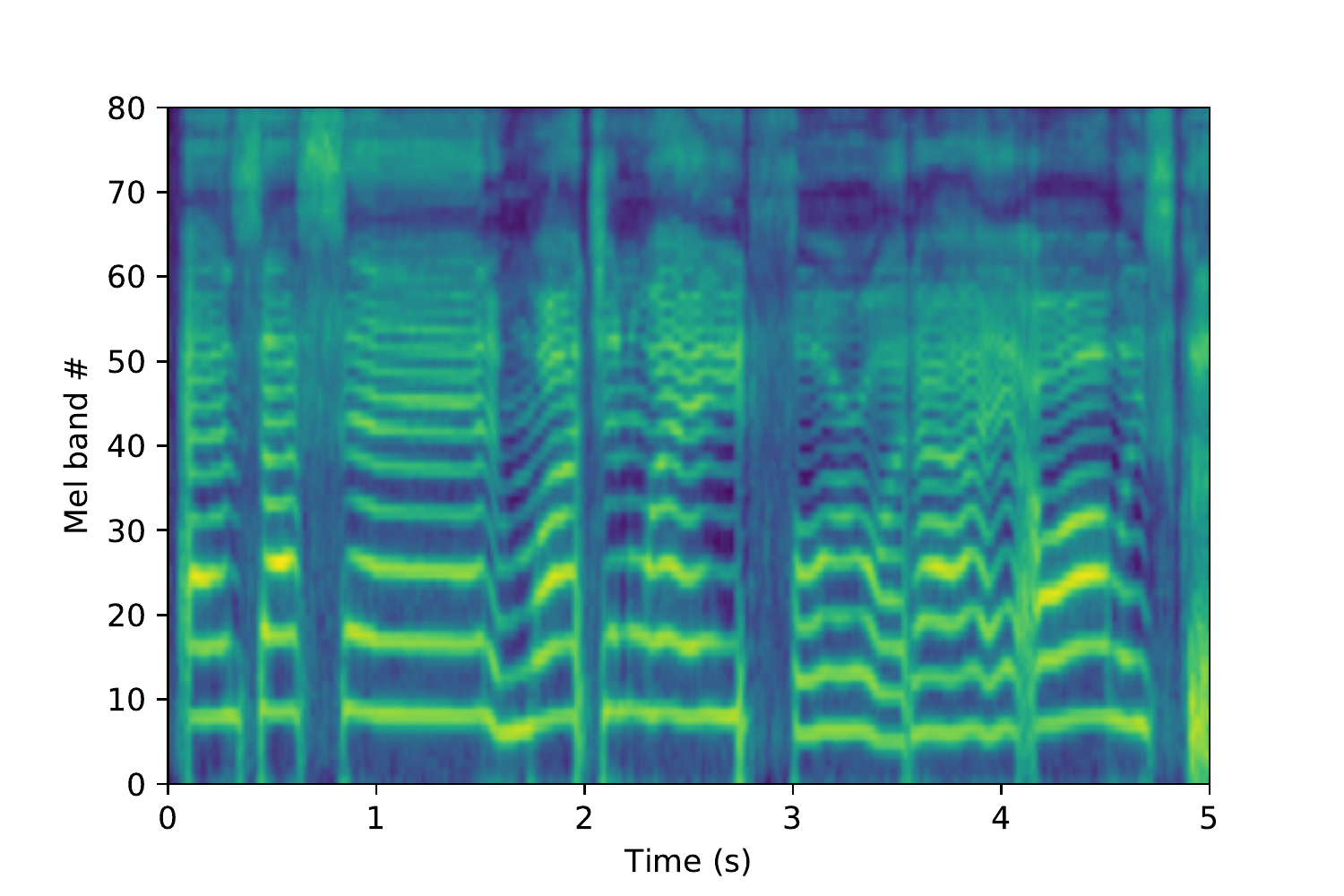}}
	\vspace{0.25em}(b)
  \end{minipage}
  \begin{minipage}{.33\textwidth}
	\centering
  \centerline{\includegraphics[width=0.95\columnwidth]{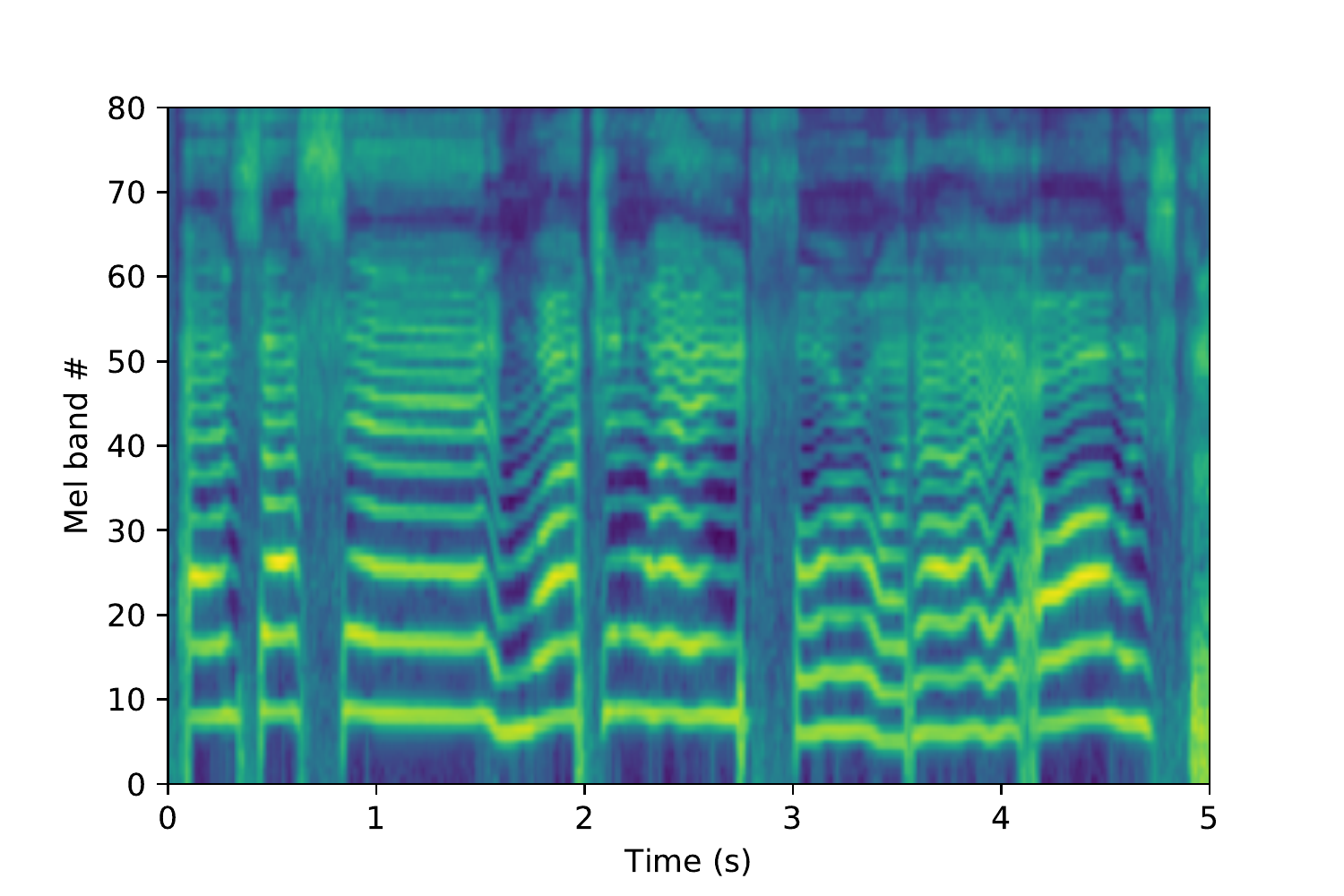}}
	\vspace{0.25em}(c)
  \end{minipage}

\begin{minipage}{.33\textwidth}
	\centering
  \centerline{\includegraphics[width=0.95\columnwidth]{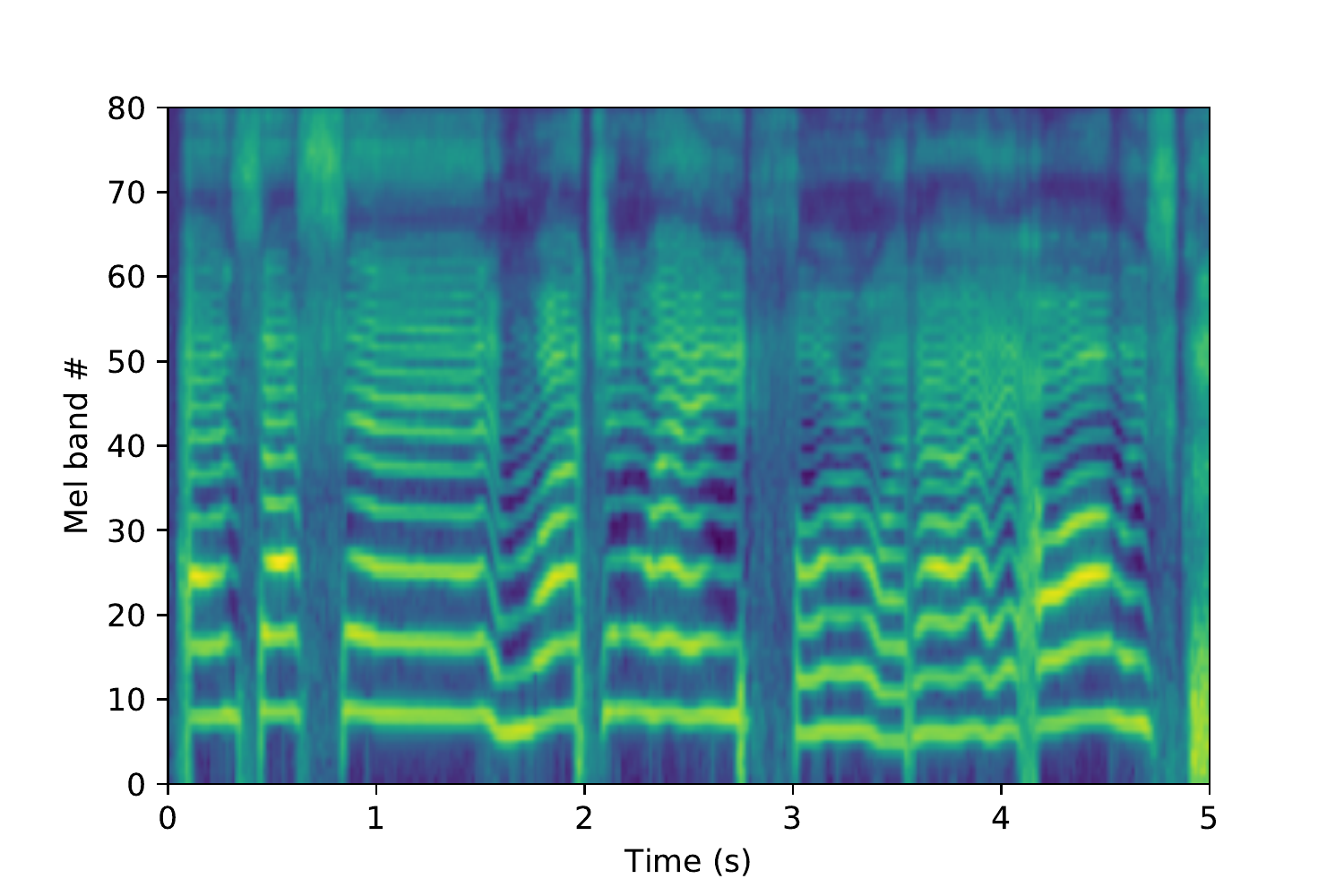}}
	\vspace{0.25em}(d)
  \end{minipage}
  \begin{minipage}{.33\textwidth}
	\centering
  \centerline{\includegraphics[width=0.95\columnwidth]{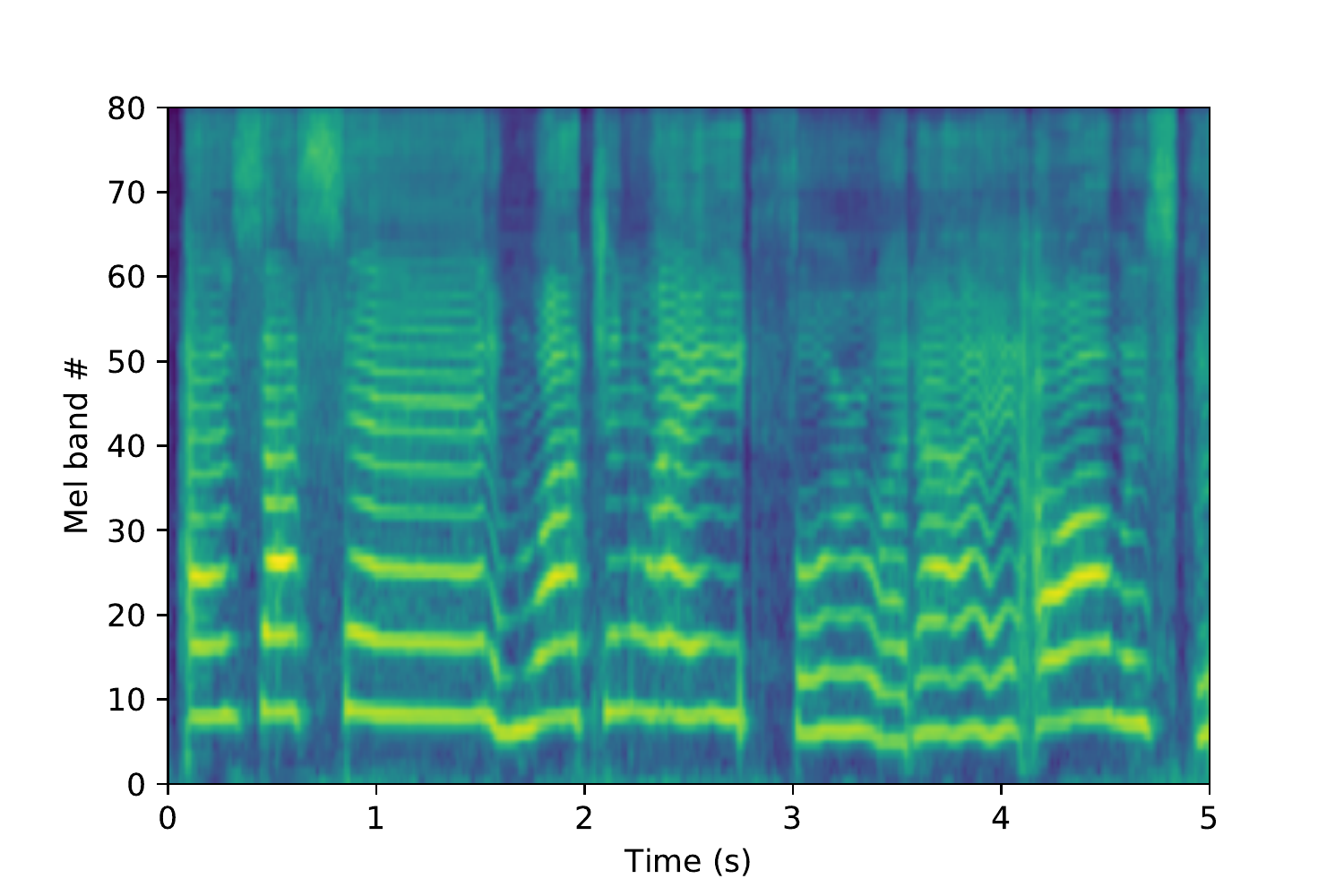}}
	\vspace{0.25em}(e)
  \end{minipage}
  \begin{minipage}{.33\textwidth}
	\centering
  \centerline{\includegraphics[width=0.95\columnwidth]{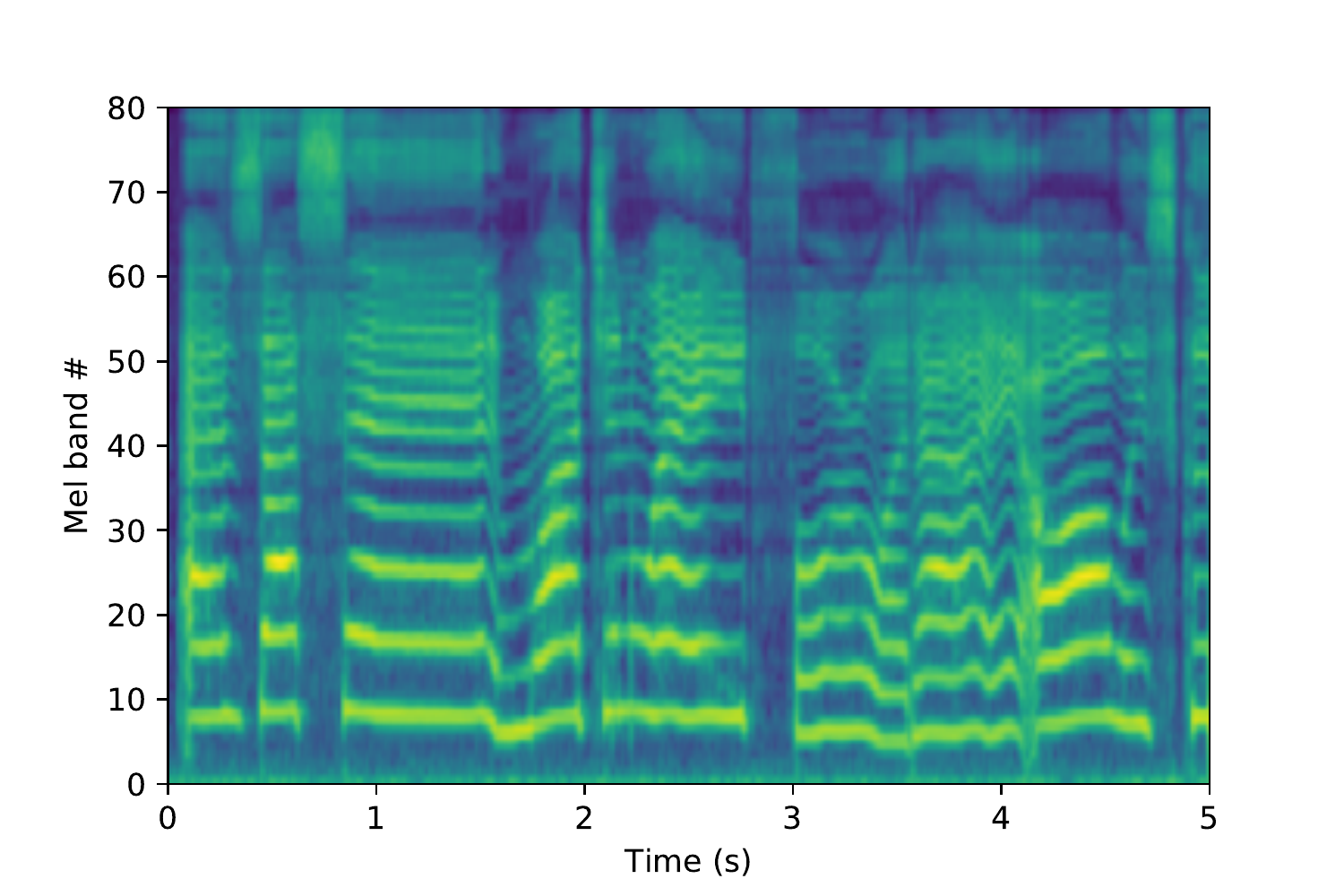}}
	\vspace{0.25em}(f)
  \end{minipage}

\caption{\it Log Mel spectrogram of (a) original signal, inversion of WORLD acoustic features using (b) non-differentiable WORLD synthesizer, (c) differentiable counterpart, (d) with added compression/decompression, (e) with added trainable postnet, and (f) using the alternate excitation method.}
\label{fig:inversion}
\end{figure*}

\subsection{End-to-end audio style transfer}
We train several different models to (non-exhaustively) exemplify the proposed methodologies.  To this end, we focus our attention on the training of audio style transfer systems in an end-to-end manner, as we have established its advantages over component-level training in our previous work \cite{Shahan3}.  The SVC task is trained on approximately 90 minutes of proprietary singing voice data of a single target singer.  Similarly, the DDSP-TT task is trained on approximately 20 minutes of proprietary violin data of a single performer.  All systems are trained at a sampling rate of 22050 Hz, using 2-second audio clips and a batch size of 4.  We use the ADAM optimizer with a learning rate of $10^{-4}$ and train for 1M training steps.  When adversarial training is enabled, we train the generator for 50K steps without including adversarial loss terms or training of the discriminator.

For the SVC task, we train four variants of our system using different combinations of proposed vocoders and objective function:  differentiable WORLD synthesizer (E2E+DiffWORLDSynth), E2E+DiffWORLDSynth with postnet and adversarial training (E2E+DiffWORLDSynth+Postnet+GAN), E2E+DiffWORLDSynth with constrained-manifold targets (E2E+DiffWORLDSynth+Oracle), and the alternate differentiable WORLD source excitation method (E2E+DiffWORLDExcite).  For the DDSP-TT task, we simply consider the analogous flavor of E2E+DiffWORLDSynth.  The WORLD log Mel spectrogram representation and its associated differentiable decompression is assumed in all systems.

In order to get a notional sense of system performance, we simply compute an L1 reconstruction error of log Mel spectrograms computed between inferred waveform reconstructions and their respective system targets, as illustrated in Table \ref{tab:quant1}. To this end, E2E+DiffWORLDSynth serves as a baseline system for comparison. Interestingly, the addition of postnet and adversarial loss terms in E2E+DiffWORLDSynth+Postnet+GAN results in slightly worse average performance, though it can improve plausibility on any one given output, as illustrated in Figure \ref{fig:mle_gan}.  Specifically, we see that networks trained with added adversarial loss terms in their objective function may synthesize spectral envelopes with finer details and tighter formant bandwidths (which can reduce buzziness).  While E2E+DiffWORLDSynth+Oracle uses the same architecture as that in E2E+DiffWORLDSynth, its error metric is significantly lower because in this case, we compare reconstructions against WORLD synthesized outputs assuming ground truth 
\begin{table}[hb]
\centering
  \begin{tabular}{clc}
    \toprule
Task & System & Error \\ \midrule
SVC & E2E+DiffWORLDSynth & {0.2050}     \\
SVC & E2E+DiffWORLDSynth+Postnet+GAN & {0.2215}   \\
SVC & E2E+DiffWORLDSynth+Oracle & {0.1014}  \\
SVC & E2E+DiffWORLDExcite & \bf{0.0421}  \\  \midrule
DDSP-TT & E2E+DiffWORLDSynth & \bf{0.2465}  \\ \bottomrule
  \end{tabular}
 \caption{Mel reconstruction losses for various audio style transfer systems on their respective test sets.}
 \label{tab:quant1}
\end{table}
\begin{figure*}[ht]
  \begin{minipage}{.49\textwidth}
	\centering
  \centerline{\includegraphics[width=0.95\columnwidth]{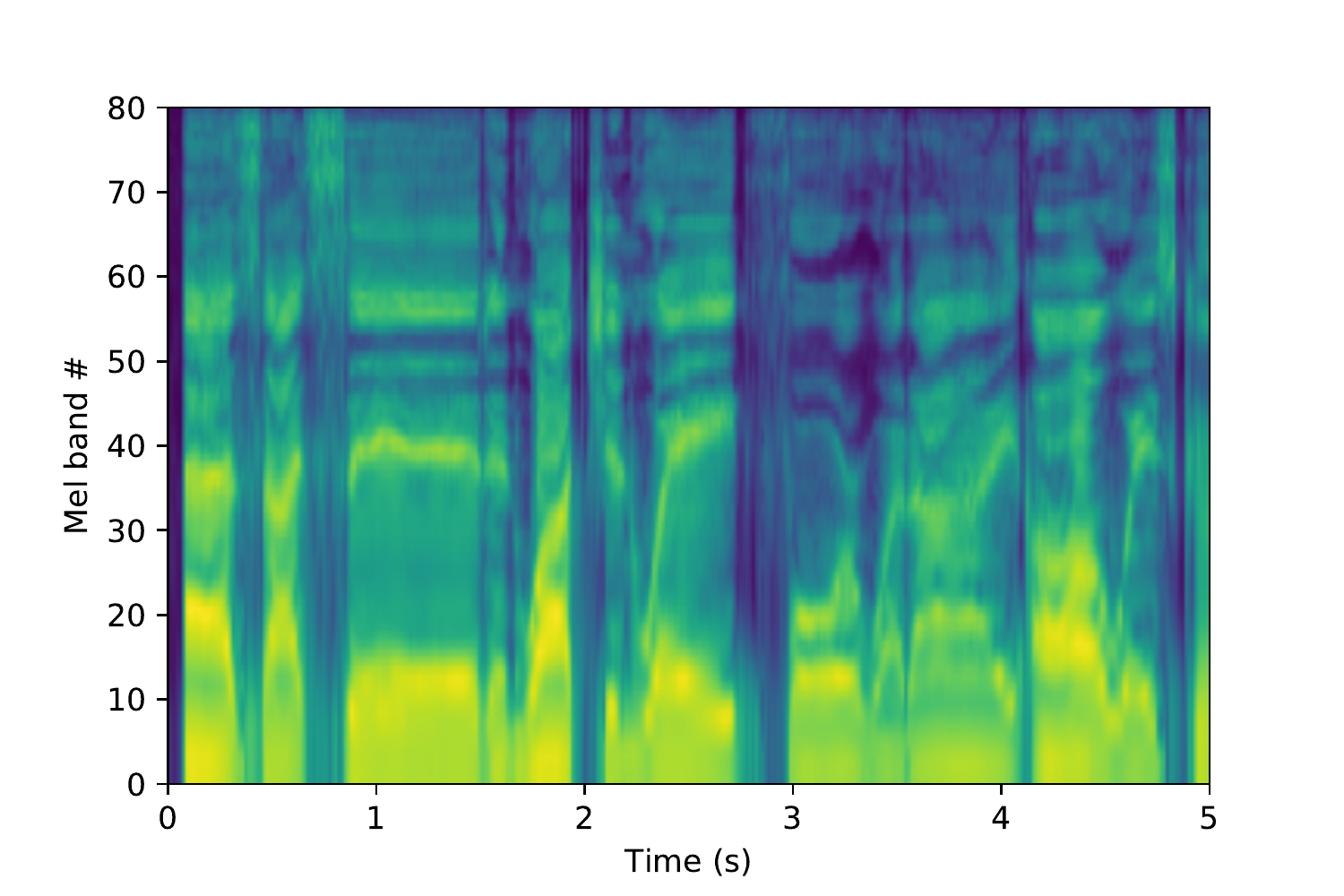}}
	\vspace{0.25em}(a)
  \end{minipage}
  \begin{minipage}{.49\textwidth}
	\centering
  \centerline{\includegraphics[width=0.95\columnwidth]{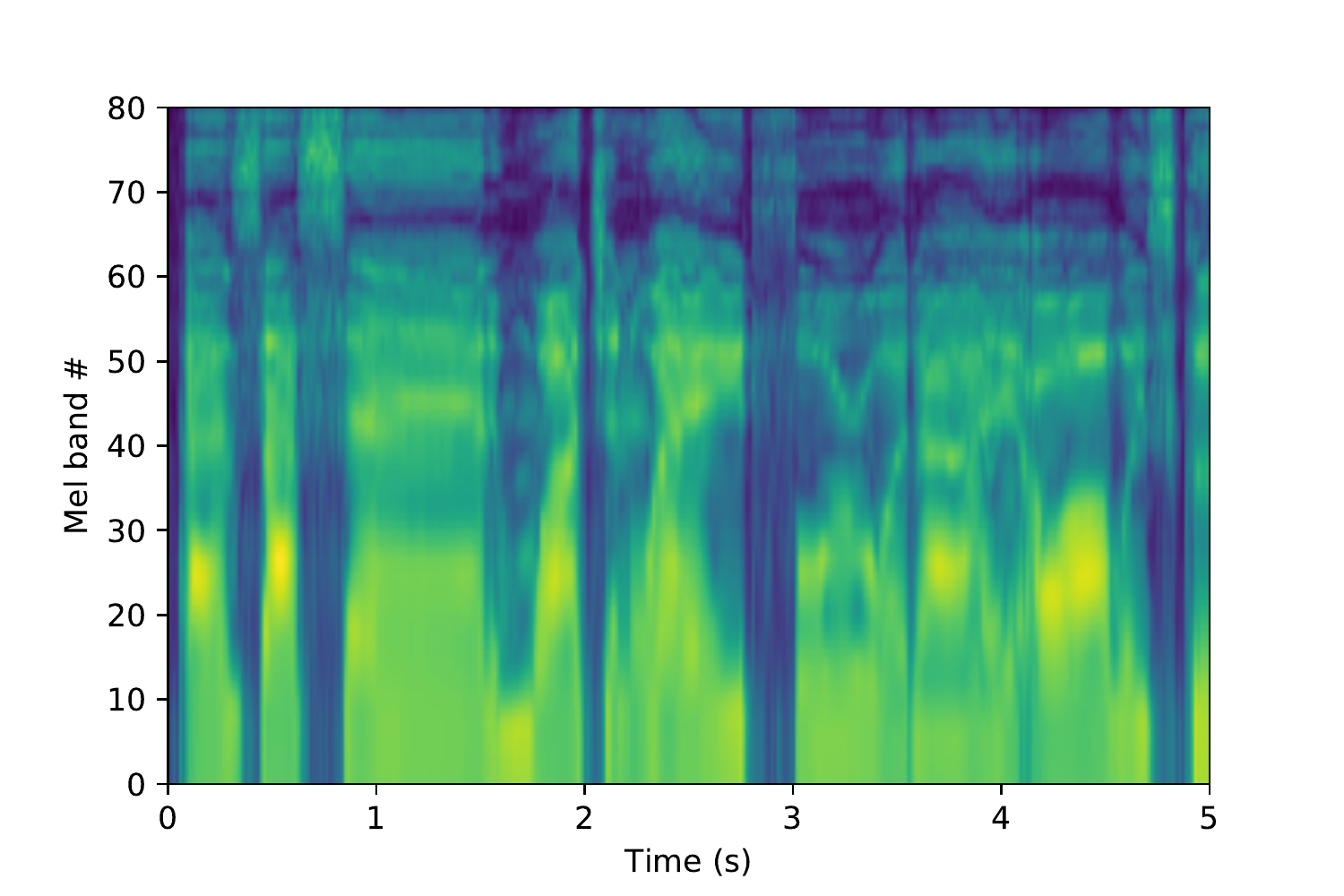}}
	\vspace{0.25em}(b)
  \end{minipage}
  \begin{minipage}{.49\textwidth}
	\centering
  \centerline{\includegraphics[width=0.95\columnwidth]{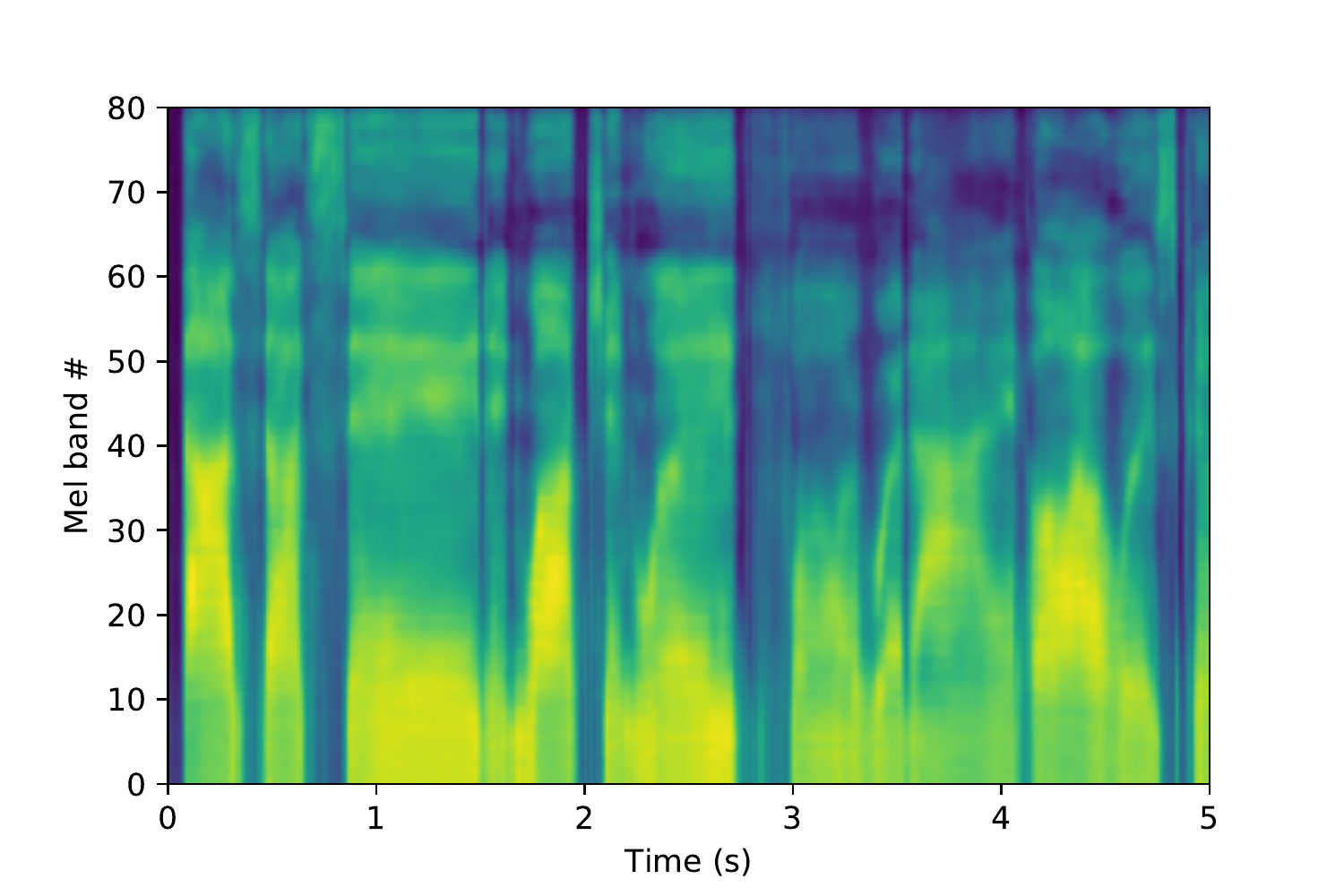}}
	\vspace{0.25em}(e)
  \end{minipage}
  \begin{minipage}{.49\textwidth}
	\centering
  \centerline{\includegraphics[width=0.95\columnwidth]{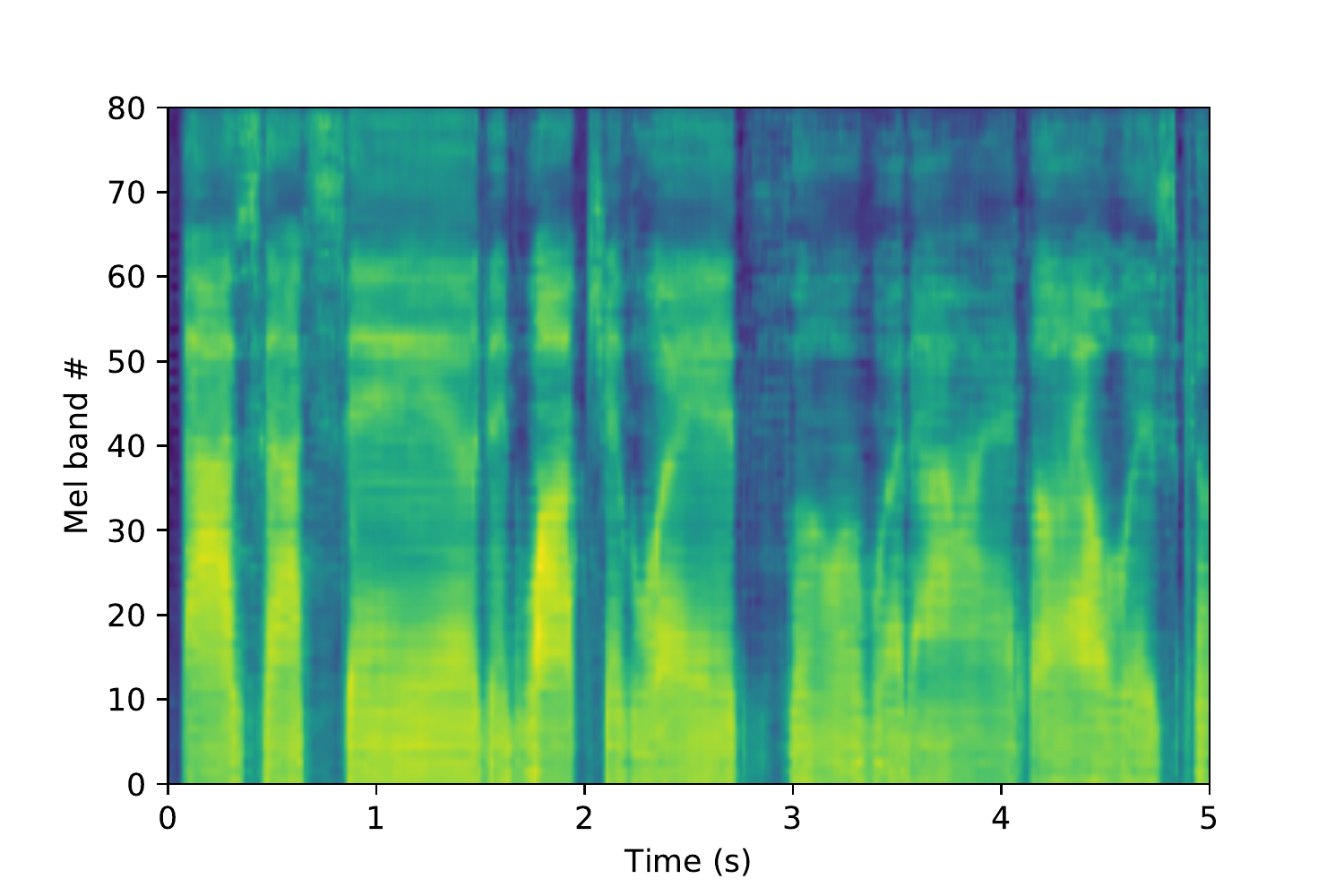}}
	\vspace{0.25em}(f)
  \end{minipage}

\caption{\it WORLD log Mel spectrogram of (a) source signal, (b) target signal, (c) inferred by a network trained (c) without adversarial loss terms, and (d) with added adversarial loss terms in their objective functions.  Note that (a) and (b) reflect parallel examples for illustrative purposes, but parallel data is not a requirement for system training.}
\label{fig:mle_gan}
\end{figure*}
acoustic parameters (i.e. not against the original audio).  As is to be expected, E2E+DiffWORLDExcite appears to be the best performer according to the reconstruction error metric, as it allows for an identity operation given access to a suitable excitation spectrum.  However, this system on its own does not provide voice conversion completely.  In practice, it would need to rely on an accurate pitch-shifting procedure in order to provide anything other than a formant transformation.  Lastly, the E2E+DiffWORLDSynth is capable of performing the DDSP-TT task, with the benefit that it was possible to add an acoustic feature parameter loss to its objective function.

\section{Conclusions}
\label{sec:conclusions}

We presented a new class of neural vocoders based on a differentiable implementation of the WORLD synthesizer, and applied them to different end-to-end audio style transfer pipelines.  Experimental results illustrated that indeed, our differentiable implementation of the synthesizer mirrors closely to its non-differentiable predecessor.  The different end-to-end training methodologies that it enables and was outlined in this work are found to be tractable.  Future research will leverage this work in order to develop lightweight, low-latency SVC systems. Furthermore, we would like to exploit further properties of our differentiable synthesizer in order to enable new layer types for audio manipulation and novel audio style transfer solutions thereof.

\bibliographystyle{plain}
\bibliography{refs}

\begin{thebibliography}{10}

\bibitem{SingingSynthesis}
M.~Blaauw and J.~Bonada.
\newblock A neural parametric singing synthesizer.
\newblock In {\em Proc. of Interspeech}, 2017.

\bibitem{WganSing}
P.~Chandna, M.~Blaauw, J.~Bonada, and E.~Gomez.
\newblock {WGANSing}: A multi-voice singing voice synthesizer based on the
  {Wasserstein-GAN}.
\newblock In {\em Proc. of the 27th European Signal Processing Conference},
  2019.

\bibitem{Ddsp}
J.~Engel, L.~Hantrakul, C.~Gu, and A.~Roberts.
\newblock {DDSP}: Differentiable digital signal processing.
\newblock In {\em Proc. of the International Conference on Learning
  Representations}, pages 26--30, 2020.

\bibitem{Ddspeech}
G.~Fabbro, V.~Golkov, T.~Kemp, and D.~Cremers.
\newblock Speech synthesis and control using differentiable {DSP}.
\newblock {\em arXiv:2010.15084}, 2020.

\bibitem{Timit}
J.~S. Garapolo et~al.
\newblock {\em {TIMIT} Acoustic-Phonetic Continuous Speech Corpus {LDC93S1}}.
\newblock Linguistic Data Consortium, Philadelphia, 1993.

\bibitem{WaveRnn}
N.~Kalchbrenner et~al.
\newblock Efficient neural audio synthesis.
\newblock {\em arXiv:1802.08435}, 2018.

\bibitem{MelGAN}
K.~Kumar, R.~Kumar, T.~de~Boissiere, L.~Gestin, W.Z. Teoh, J.~Sotelo,
  A.~de~Br\'{e}bisson, Y.~Bengio, and A.C. Courville.
\newblock Melgan: Generative adversarial networks for conditional waveform
  synthesis.
\newblock In {\em Advances in Neural Information Processing Systems},
  volume~32, 2019.

\bibitem{Lent}
K.~Lent.
\newblock An efficient method for pitch shifting digitally sampled sounds.
\newblock {\em Computer Music Journal}, 13(4):65--71, 1989.

\bibitem{SeaNet}
Y.~Li, M.~Tagliasacchi, O.~Rybakov, V.~Ungureanu, and D.~Roblek.
\newblock Real-time speech frequency bandwidth extension.
\newblock In {\em Proc. of the IEEE International Conference on Acoustics,
  Speech and Signal Processing (ICASSP)}, pages 691--695, 2021.

\bibitem{HooliGan}
O.~{McCarthy} and Z.~Ahmed.
\newblock Hooli{GAN}: Robust, high quality neural vocoding.
\newblock {\em arXiv:2008.02493}, 2020.

\bibitem{WORLD}
M.~Morise, F.~Yokomori, and K.~Ozawa.
\newblock {WORLD}: a vocoder-based high-quality speech synthesis system for
  real-time applications.
\newblock {\em IEICE Transactions on Information and Systems},
  E99-D(7):1877--1884, 2016.

\bibitem{Shahan1}
S.~Nercessian.
\newblock Improved zero-shot voice conversion using explicit conditioning
  signals.
\newblock In {\em Proc. of Interspeech}, pages 4711--4715, 2020.

\bibitem{Shahan2}
S.~Nercessian.
\newblock Zero-shot singing voice conversion.
\newblock In {\em Proc. of the International Society for Music Information
  Retrieval Conference}, page 70–76, 2020.

\bibitem{Shahan3}
S.~Nercessian.
\newblock End-to-end zero-shot voice conversion using a {DDSP} vocoder.
\newblock In {\em Proc. of IEEE Workshop on Applications of Signal Processing
  to Audio and Acoustics (WASPAA)}, pages 1--5, 2021.

\bibitem{Tacotron}
J.~Shen et~al.
\newblock Natural {TTS} synthesis by conditioning {WaveNet} on {Mel}
  spectrogram predictions.
\newblock In {\em Proc. of the IEEE International Conference on Acoustics,
  Speech and Signal Processing}, pages 4779--4783, 2018.

\bibitem{HifiGan2}
J.~Su, Z.~Jin, and A.~Finkelstein.
\newblock {HiFi-GAN}-2: Studio-quality speech enhancement via generative
  adversarial networks conditioned on acoustic features.
\newblock In {\em Proc. of the IEEE Workshop on Applications of Signal
  Processing to Audio and Acoustics (WASPAA)}, pages 166--170, 2021.

\bibitem{HifiGan}
J.~Su, Y.~Wang, A.~Finkelstein, and Z.~Jin.
\newblock Bandwidth extension is all you need.
\newblock In {\em Proc. of the IEEE International Conference on Acoustics,
  Speech and Signal Processing (ICASSP)}, pages 696--700, 2021.

\bibitem{LPCNet}
J.M. Vali and J.~Skoglund.
\newblock {LPCNET}: Improving neural speech synthesis through linear
  prediction.
\newblock In {\em Proc. of the IEEE International Conference on Acoustics,
  Speech and Signal Processing (ICASSP)}, pages 5891--5895, 2019.

\bibitem{WaveNet}
A.~van~den Oord et~al.
\newblock {WaveNet}: A generative model for raw audio.
\newblock {\em arXiv:1609.03499}, 2016.

\bibitem{Nsf2}
X.~Wang and J.~Yamagishi.
\newblock {Using Cyclic Noise as the Source Signal for Neural
  Source-Filter-Based Speech Waveform Model}.
\newblock In {\em Proc. of Interspeech}, pages 1992--1996, 2020.

\end{thebibliography}

\end{sloppy}
\end{document}